\documentclass[twocolumn,trackchanges]{aastex7}
\usepackage[T1]{fontenc}
\usepackage{stackengine}
\usepackage{amsmath}
\usepackage{placeins}

\newcommand{\Msun}{M_{\odot}}
\newcommand{\MHI}{M_{\rm H\,{\textsc i}}}
\newcommand{\Mstar}{M_{*}}
\newcommand{\fHI}{f_{\rm H\,{\textsc i}}}
\newcommand{\Lfoursp}{L_{144}}
\shorttitle{H\,{\sc i} Recovery in FASHI and LoTSS}
\shortauthors{Bera et al.}
\begin{document}

\title{H\,{\sc i} Emission Recovery in the Low-Redshift Radio Population:
Host-Galaxy and Radio-Luminosity Dependence in FASHI and LoTSS}

\correspondingauthor{Taotao Fang}
\email{fangt@xmu.edu.cn}

\author[orcid=0000-0001-7121-4258,gname='Soumen Kumar',sname='Bera']{Soumen Kumar Bera}
\affiliation{Department of Astronomy, Xiamen University, Xiamen, Fujian 361005, People's Republic of China}
\email{beras@xmu.edu.cn}
\author[orcid=0000-0002-2853-3808,gname='Taotao',sname='Fang']{Taotao Fang}
\affiliation{Department of Astronomy, Xiamen University, Xiamen, Fujian 361005, People's Republic of China}
\email{fangt@xmu.edu.cn}
\author[orcid=0000-0003-1761-5442,gname='Chuan',sname='He']{Chuan He}
\affiliation{Department of Astronomy, Xiamen University, Xiamen, Fujian 361005, People's Republic of China}
\email{hechuan@xmu.edu.cn}
\author[orcid=0000-0002-3940-2950,gname='Shulan',sname='Yan']{Shulan Yan}
\affiliation{Department of Astronomy, Xiamen University, Xiamen, Fujian 361005, People's Republic of China}
\email{yansl@stu.xmu.edu.cn}

\begin{abstract}
Overlapping blind 21\,cm and wide-area low-frequency continuum surveys make it possible to measure which radio sources an untargeted H\,{\sc i} survey recovers. Cross-matching FASHI DR1 with the LoTSS DR2 value-added catalog yields 4392 one-to-one associations. Within an empirically defined comparison region of $2912$\,deg$^2$, we construct a low-redshift, spectroscopic, radio-selected parent of 25\,573 LoTSS sources whose expected H\,{\sc i} line falls in the FASHI band, and measure how often each is recovered. Sources classified as radio-loud AGN in the published LoTSS DR2 radio-AGN classification are recovered less often than star-forming ones: 143 of 1542 ($9.27\%$, $7.87$--$10.83$) against $14.88\%$ ($14.43$--$15.33$) of 24\,031, an odds ratio of $0.585$ ($0.491$--$0.697$). The two classes, however, differ strongly in host properties, by $1.34$\,dex in median specific star formation rate, and within their overlapping stellar-mass and star-formation ranges we detect no statistically significant residual class difference (adjusted odds ratio $1.60$, $0.87$--$2.94$). In contrast, 144\,MHz luminosity remains positively associated with H\,{\sc i} recovery after adjustment for the measured host properties, redshift and an approximate sensitivity proxy: within the same overlapping range, each dex raises the odds of recovery by a factor of $2.03$ ($1.54$--$2.68$), an association carried primarily by the star-forming population, not by the radio-AGN subsample. We interpret this as an empirical relationship with detectable H\,{\sc i} emission, not as evidence that radio activity increases or decreases H\,{\sc i} mass. The association catalog and the controlled recovery measurements characterize the radio population sampled by a blind H\,{\sc i} survey.
\end{abstract}


\section{Introduction}
\label{sec:intro}
Radio jets launched by accreting supermassive black holes deposit energy into their host galaxies and into the surrounding intracluster and intergalactic medium, and are widely held to regulate the cooling of gas in massive halos \citep{2012ARA&A..50..455F, 2014ARA&A..52..589H, 2020NewAR..8801539H}. The atomic phase is a large-scale reservoir that can replenish the molecular gas from which stars form, and is the cold phase most directly accessible in emission at low redshift \citep{2022ARA&A..60..319S}. How the two are related --- whether radio activity and its observable structure are connected to the cold gas content of the host --- remains only partly settled.

The classical picture associates powerful radio galaxies with massive, quiescent early-type hosts that are poor in cold gas \citep{2005MNRAS.362...25B, 2012MNRAS.421.1569B, 2019A&A...622A..17S}. Targeted 21\,cm observations of nearby radio galaxies have complicated this picture in an instructive way. \citet{2007A&A...464L...1E, 2010MNRAS.406..987E} imaged a complete sample of nearby low-power radio galaxies and found large-scale, regularly rotating H\,{\sc i} disks and rings around roughly half of the hosts of \emph{compact} radio sources, while none of the extended FR\,I sources in the same sample showed comparable H\,{\sc i} in emission. Whatever the underlying cause --- confinement of a young radio source by a dense interstellar medium, or an accretion history that stalls before the source can expand --- the empirical association runs opposite to the intuition that larger or morphologically more complex radio sources should sit in more gas-rich hosts. Complementary constraints come from 21\,cm absorption, which traces gas on much smaller scales and frequently reveals disturbed kinematics and outflows in compact and young sources \citep{2018A&ARv..26....4M, 2017A&A...604A..43M}.

These studies share two limitations. They are based on modest samples, and their parent samples are selected on radio or optical properties, so the H\,{\sc i} information is obtained by follow-up rather than by survey. Blind H\,{\sc i} surveys invert that logic, and it is now possible to ask the question the other way round: not what the H\,{\sc i} content of a radio-selected sample is, but what part of the radio-source population a blind 21\,cm survey recovers in the first place. We therefore quantify the low-frequency radio population associated with blind H\,{\sc i} detections, and the host and radio properties that govern whether a radio source is recovered by such a survey.

The data to answer it have only recently become available. The first data release of the FAST All Sky H\,{\sc i} survey \citep[FASHI;][]{2024SCPMA..6719511Z} covers more than 7600\,deg$^2$ and catalogs 41,741 extragalactic H\,{\sc i} sources at $z \lesssim 0.09$, with a median detection sensitivity near $0.76$\,mJy\,beam$^{-1}$ and a velocity resolution of $\sim 6.4$\,km\,s$^{-1}$. Over a large fraction of the same sky, the LOFAR Two-metre Sky Survey \citep[LoTSS;][]{2022A&A...659A...1S} provides $6''$ imaging at 144\,MHz, and its Data Release 2 value-added catalog \citep{2023A&A...678A.151H} supplies optical and infrared identifications, redshifts, stellar masses, and structural information for several million radio sources. Radio AGN have been separated from star-forming galaxies across that catalog by \citet{2025MNRAS.539.1856H}, providing a published reference classification against which any subsample can be compared.

This paper asks one question: which low-redshift radio sources are recovered by a blind H\,{\sc i} survey, and whether radio luminosity carries information about that recovery beyond the measured host-galaxy properties. We address it in three steps. First we measure the raw H\,{\sc i} recovery fraction of radio sources carrying a catalog radio-AGN classification against that of the star-forming remainder, within a radio-selected parent drawn from one sky area, redshift range and flux limit. Second we compare the two classes within the range of host properties they share, controlling for stellar mass and specific star formation rate, on which atomic gas content is already known to depend, and for redshift and survey sensitivity. Third we test whether adding 144\,MHz luminosity to the same host-controlled model improves its description of recovery. Throughout we distinguish H\,{\sc i} \emph{recovery} --- whether a radio source is matched to a FASHI detection --- from the intrinsic atomic gas content of the host, which recovery alone cannot constrain.

Several selection effects require explicit control. Atomic gas fraction falls systematically with stellar mass in the local galaxy population \citep{2010MNRAS.403..683C, 2018MNRAS.476..875C}, and H\,{\sc i} comparisons between AGN hosts and other galaxies depend on matching in star formation rate as well as stellar mass \citep{2019MNRAS.482.5694E}, so a difference in recovery between two classes that occupy different host-property ranges may be inherited from those scalings rather than caused by whatever the classes are taken to measure. The radio-AGN classification we adopt is a cut in a plane defined by radio luminosity and $W3$ magnitude, so it depends on the depth and quality of the $W3$ photometry as much as on the radio data. The LoTSS component code, which we examine separately in Appendix~\ref{app:regression}, depends on angular resolution and surface brightness and therefore on distance \citep{2019MNRAS.488.2701M, 2023A&A...678A.151H}. A blind 21\,cm survey is subject to a flux limit that rises steeply with distance, imposing a Malmquist-type bias in H\,{\sc i} mass \citep{2011AJ....142..170H, 2018ApJ...861...49H}. Separating accretion-powered from star-formation-powered radio emission is itself difficult at the low luminosities a blind H\,{\sc i} survey reaches, where the two contribute comparably and where the standard diagnostics are calibrated on brighter samples \citep{2018MNRAS.475.3010G, 2021A&A...648A...6S}. Finally, FAST has a beam of $\sim 2.9'$ FWHM \citep{2019SCPMA..6259502J}, about thirty times coarser than LoTSS, so a single H\,{\sc i} detection may in principle include emission from more than one galaxy \citep{2015MNRAS.449.1856J, 2016MNRAS.457.4393J}.

An equally important limitation is what a non-detection means. Requiring an H\,{\sc i} emission detection selects against precisely the gas-poor early-type hosts that dominate classical radio-galaxy samples, so the recovered subset is not representative of the radio population; because our parent is radio-selected rather than H\,{\sc i}-selected, however, the quantity we measure is the recovery fraction of that population rather than the composition of an H\,{\sc i} sample. Nothing in this paper constrains the atomic gas content of radio galaxies as a class, since a global 21\,cm non-detection cannot distinguish an intrinsically H\,{\sc i}-poor host from one whose H\,{\sc i} mass lies below the FASHI limit at its distance.

Previous studies have measured atomic gas in targeted radio-galaxy samples \citep{2007A&A...464L...1E, 2010MNRAS.406..987E} and in optically selected AGN hosts \citep{2019MNRAS.482.5694E, 2026arXiv260625367S}. Here we instead use a low-redshift, radio-selected LoTSS parent sample and independently identified FASHI H\,{\sc i}-emission sources to quantify which radio sources are recovered by a blind H\,{\sc i} survey. Beyond measuring raw recovery fractions for the nominal radio-AGN and star-forming classes, we test whether radio luminosity retains predictive information once measured host-galaxy properties and observational selection are controlled. This provides a new empirical characterization of the radio population accessible to H\,{\sc i}-emission surveys, without requiring a causal interpretation in terms of fueling or feedback. Our central result is that, at fixed measured stellar mass, specific star formation rate, redshift and survey sensitivity, 144\,MHz luminosity is positively associated with H\,{\sc i} recovery in a population dominated by star-forming sources, while no residual difference between the nominal classes is statistically established where their host properties overlap. The FASHI/LoTSS association catalog is released alongside the paper.

Section~\ref{sec:data} describes the catalogs, the cross-match, the comparison region and sensitivity treatment, the H25 classification, and the host stellar masses and star formation rates. Section~\ref{sec:results} presents the analysis: the raw H\,{\sc i} recovery fractions (Section~\ref{subsec:rates}), the comparison at fixed host properties (Section~\ref{subsec:hostcontrol}), the dependence on 144\,MHz luminosity (Section~\ref{subsec:lum}), the robustness checks that bear on both (Section~\ref{subsec:robust}), recovery across the $\Lfoursp$--$M_{W3}$ plane (Section~\ref{subsec:surface}), and association quality and beam confusion (Section~\ref{subsec:association}). The LoTSS component code is examined in Appendix~\ref{app:regression} and mid-infrared host colors in Appendix~\ref{app:wise}. Section~\ref{sec:discussion} discusses what these results do and do not establish, together with the limitations of the analysis, and Section~\ref{sec:conclusions} summarizes. We adopt a flat $\Lambda$CDM cosmology with $H_0 = 70$\,km\,s$^{-1}$\,Mpc$^{-1}$, $\Omega_m = 0.3$ and $\Omega_\Lambda = 0.7$, matching the cosmology used to derive the luminosities and stellar masses in the LoTSS DR2 value-added catalog \citep{2023A&A...678A.151H} and in \citet{2025MNRAS.539.1856H}.

\section{The Data Sample}
\label{sec:data}

\subsection{Parent Catalogs}
Our H\,{\sc i} information comes from the first FASHI data release \citep{2024SCPMA..6719511Z}. FASHI is a drift-scan survey with the Five-hundred-meter Aperture Spherical radio Telescope (FAST), designed to cover the declination range $-14^\circ$ to $+66^\circ$. The first release covers more than 7600\,deg$^2$ and contains 41,741 extragalactic H\,{\sc i} sources in the frequency range 1305.5--1419.5\,MHz, corresponding to $z \lesssim 0.09$. For each source we take the H\,{\sc i} centroid position and heliocentric velocity, the line widths at 50\% and 20\% of the peak ($W_{50}$, $W_{20}$), the tabulated rms noise and signal-to-noise ratio, the adopted distance, and the H\,{\sc i} mass. The public data release comprises the source catalog and four auxiliary cross-identification tables; it includes no coverage, exposure or noise product, and the quoted declination range of $-14^\circ$ to $+66^\circ$ is the survey design rather than the DR1 footprint. We therefore bound the usable sky area from the detections themselves (Section~\ref{subsec:parent}), adopt the published searchable frequency range as the usable spectral band, and model the local sensitivity from the per-source noise (Section~\ref{subsec:sens}).

Radio continuum and host information come from the LoTSS DR2 value-added catalog \citep{2023A&A...678A.151H}, built on the $6''$ 144\,MHz imaging of \citet{2022A&A...659A...1S}. The catalog provides optical and infrared counterparts drawn from the DESI Legacy Imaging Surveys and unWISE, redshifts (spectroscopic where available, otherwise photometric; \citealt{2022MNRAS.512.3662D}), SED-derived stellar masses with an associated reliability flag, rest-frame photometry, 144\,MHz total flux densities and $k$-corrected luminosities $\Lfoursp$ (computed for $\alpha=0.7$), a resolution flag and largest angular size for resolved sources, and the source structure code \texttt{S\_Code}.

We use principally the 144\,MHz total flux density and $k$-corrected luminosity, the host redshift and its origin, the SED stellar mass and its reliability flag, the component code \texttt{S\_Code}, and the Legacy and unWISE magnitudes with their uncertainties; the complete field list is given in the machine-readable table description.

\subsection{Cross-matching and Sample Definition}
\label{subsec:xmatch}
The FAST beam is far larger than the LoTSS synthesized beam, so a positional match alone is not sufficient to establish a physical association. We paired each FASHI H\,{\sc i} centroid with its nearest LoTSS optical counterpart position within $90''$ --- approximately the half-power radius of the FAST beam, $\approx 87''$ --- keeping only mutual nearest neighbours, so that the detection is also the FASHI source nearest to that counterpart, and then required a velocity difference
\begin{equation}
\Delta v = \left| cz_{\rm H\,\textsc{i}} - c\,z_{\rm best} \right|
           < 500\ {\rm km\,s^{-1}},
\label{eq:vdiff}
\end{equation}
where $cz_{\rm H\,\textsc{i}}$ is the FASHI heliocentric velocity in the optical convention and $z_{\rm best}$ is the host redshift from the LoTSS DR2 value-added catalog. We note that the FASHI catalog also lists a radio-convention velocity, $v_{\rm radio}$, for the same H\,{\sc i} line; it is not a host systemic velocity and is not used here. We verified the tabulated separations independently by reconstructing the FASHI positions from the cataloged source names; the two agree to within $0.27''$ for every source. This yielded 4392 accepted associations.

The accepted table is one to one: it contains 4392 unique FASHI identifiers and 4392 unique LoTSS sources, so no FASHI H\,{\sc i} measurement is assigned to more than one radio host and no host appears more than once. The one-to-one assignment is a convention rather than a measurement, since a single-dish detection cannot be apportioned between galaxies, and because positions are paired before velocities are compared, a detection whose nearest optical counterpart fails Equation~(\ref{eq:vdiff}) is assigned to no host even when another counterpart within $90''$ satisfies it. Within the comparison parent of Section~\ref{subsec:parent}, 829 sources ($3.2$ per cent) lie within $90''$ and $500$\,km\,s$^{-1}$ of a FASHI detection without being credited with it, at a median separation of $48''$ against $21''$ for the accepted associations. Of these, 631 are themselves the nearest velocity-consistent counterpart of a detection that the positional step did not pair with them, 152 lie near a detection already credited to a closer parent source, and 46 lie near a detection whose nearest velocity-consistent counterpart is another source. Crediting each detection instead to its nearest velocity-consistent counterpart adds the 631 and gives recovery fractions of $11.15$ and $17.38$ per cent for the radio-AGN and star-forming classes, with an odds ratio of $0.597$ ($0.507$--$0.702$) against $0.585$, so the contrast of Section~\ref{subsec:rates} does not depend on the convention; this assignment is retained as a robustness check (Section~\ref{subsec:robust}). Repeating the search while retaining every candidate, using the \citet{2025MNRAS.539.1856H} parent as the candidate list, 3749 of the 3901 detections whose accepted host lies in that list ($96.1$ per cent) have exactly one candidate, 144 have two, seven have three and one has four, and in every case the accepted host is the nearest candidate in that list, so no tie-breaking is involved. These 161 additional candidates include the 152 parent sources above; the larger total of 829 also counts sources near detections that were assigned to no host. We also estimated the chance rate by repeating the match with all FASHI positions displaced by between $0.5^\circ$ and $2^\circ$ in right ascension or declination, holding the velocities fixed so that the test measures chance positional \emph{and} kinematic coincidence. Averaged over 16 realizations this yields $26.5 \pm 7.2$ matches, or $0.60 \pm 0.16$ per cent of the 4392 accepted associations. Because that candidate list is a subset of the full value-added catalog used for the original match, the chance rate is a lower limit and the single-candidate fraction an upper limit.

Of the 4392 accepted associations, 4381 (99.7\%) have a spectroscopic host redshift (4353 SDSS, 28 DESI) and 11 have only a photometric redshift; in the mass-restricted sample of Appendix~\ref{app:regression} the
corresponding numbers are 494 and 6. This is a consequence of the velocity criterion itself: a photometric redshift uncertainty of $\sigma_z \sim 0.002$ corresponds to $\sim\!600$\,km\,s$^{-1}$, so photometric hosts rarely satisfy Equation~(\ref{eq:vdiff}). The velocity criterion therefore acts in practice as a spectroscopic-redshift requirement, and this should be understood as part of the sample definition.

We define two samples in the main text, and state at each point below which of them a quantity refers to. The \emph{accepted association sample} comprises all 4392 matches. The \emph{H25-classifiable sample} is the subset that satisfies the parent criteria of \citet{2025MNRAS.539.1856H}, hereafter H25, and therefore carries a classification. Restricted to the comparison region and usable coverage of Section~\ref{subsec:parent}, it contains 3724 sources. Membership is equivalent to requiring a total flux density above 1.1\,mJy, a host redshift above 0.01, and a $W3$ measurement, a rule we verified reproduces H25 catalog membership in both directions for all 4392 associations. Sky positions throughout are the optical-identification positions of the value-added catalog, the same positions used for the cross-match, so that the comparison region and the association are defined on one coordinate system. A third sample, restricted to hosts with a reliable stellar mass, is used only for the component-code analysis of Appendix~\ref{app:regression} and is defined there.

\begin{deluxetable*}{lccc}
\tablecaption{Sample flow from the radio-selected parent to H\,{\sc i} recovery. Classifications are those of \citet{2025MNRAS.539.1856H}: RLAGN denotes the inclusive radio-loud AGN sample (\texttt{AGN\_BROAD}) and SFG the remainder. Class counts are given from the point at which the coverage criteria have been applied.
\label{tab:selection}}
\tablehead{\colhead{Stage} & \colhead{$N$} & \colhead{RLAGN} & \colhead{SFG}}
\startdata
H25 radio-selected parent, all sky                      & 963\,764 & \nodata & \nodata \\
\quad inside the comparison rectangle                   & 600\,567 & \nodata & \nodata \\
\quad\quad $0.01 < z_{\rm best} < 0.0897$               &  28\,871 & \nodata & \nodata \\
\quad\quad\quad H\,{\sc i} line inside the usable band  &  28\,176 &    1881 & 26\,295 \\
\quad\quad\quad\quad spectroscopic host redshift        &  26\,686 &    1617 & 25\,069 \\
\quad\quad\quad\quad\quad outside flagged drift segments & 25\,573 & 1542 & 24\,031 \\
\hline
\quad\quad\quad\quad\quad\quad recovered by FASHI       &    3718 &     143 &   3575 \\
\quad\quad\quad\quad\quad\quad not recovered            &  21\,855 &    1399 & 20\,456 \\
\hline
Retaining photometric hosts (robustness)                &  26\,890 &    1787 & 25\,103 \\
\quad recovered by FASHI                                &    3724 &     144 &   3580 \\
\quad not recovered                                     &  23\,166 &    1643 & 21\,523 \\
\enddata
\tablecomments{Sky positions throughout are the optical-identification positions of the LoTSS DR2 value-added catalog, the same positions used for the cross-match. The band criterion requires the expected H\,{\sc i} line frequency, computed from $z_{\rm best}$, to fall within 1305.5--1419.5\,MHz; it is applied identically to every source irrespective of whether a FASHI counterpart exists. The drift-segment criterion removes the sources lying in the 89 segments of the rectangle flagged as potentially poorly covered based on their FASHI detection counts, those containing no detections or significantly fewer than expected (Section~\ref{subsec:parent}). The final block repeats the flow without the spectroscopic requirement, outside the same segments, and is not a further stage.}
\end{deluxetable*}

The full matched catalog of 4392 sources is provided in machine-readable form, including for each source the FASHI and LoTSS identifiers, the H\,{\sc i} and host properties used here, the angular separation and velocity difference, the \citet{2025MNRAS.539.1856H} classification, and flags recording membership of the analysis and association samples, photometric host redshifts, extreme gas fractions, near-threshold associations, and in-beam FASHI pairs.

\subsection{The Comparison Parent Sample}
\label{subsec:parent}
Comparing the composition of the associated sample with the H25 catalog as a whole would conflate H\,{\sc i} selection with differences in sky coverage, redshift and luminosity range. We therefore define a parent sample over the same observational domain as the associations. Because FASHI DR1 publishes no coverage product, that domain has to be bounded rather than measured, and we do so using two independent tracers of where both surveys observed.

The FASHI detections in DR1 span $-6.2^\circ < \delta < +65.3^\circ$, but those south of $+2^\circ$ lie outside LoTSS DR2, and within the LoTSS declination range their surface density is stable across $30^\circ < \delta < 64^\circ$ and falls to zero below $28^\circ$ and above $66^\circ$. In right ascension the number of FASHI detections per LoTSS source is approximately flat across the LoTSS DR2 13h field, $120^\circ < \alpha < 255^\circ$, and departs sharply from that level beyond it, where one survey or the other loses coverage. We therefore adopt the rectangle $30^\circ \le \delta < 64^\circ$, $120^\circ \le \alpha < 255^\circ$, a solid angle of $3085$\,deg$^2$, as the outer boundary of the comparison region. Its boundaries were fixed from these counts before any recovery rate was computed.

Two checks support the choice, and a third refines it. The LoTSS source density is uniform across the rectangle at $195$\,deg$^{-2}$ (10th to 90th percentile $182$--$212$), confirming that the radio side is complete, and the rectangle contains 18\,045 of the 41\,741 FASHI detections and 4208 of the 4392 accepted associations, or $95.8$ per cent. FASHI coverage is nonetheless not uniform within it: \citet{2024SCPMA..6719511Z} note that some sky ranges have been scanned repeatedly while others could not be covered at all. Because the survey is a drift scan with primary drift centres $21.65'$ apart in declination \citep{2024SCPMA..6719511Z}, a missing drift appears as a strip at fixed declination, which a coarse grid can miss. We therefore divided the rectangle into 1316 segments, 94 declination strips of $0.36^\circ$ (one drift spacing) by $10^\circ$ in right ascension, and compared the number of detections in each with the expectation from the neighbouring strips. Eighty-nine segments contain no detection or significantly fewer than expected ($p < 10^{-5}$); 81 of them are empty, against $1.8$ expected by chance. Together they cover $173$\,deg$^2$, $5.6$ per cent of the rectangle, and hold 20 detections and 1113 of the spectroscopic parent sources defined below, five of them recovered. We flag these segments as potentially poorly covered based on their FASHI detection counts, an empirical assessment rather than an independent confirmation of their observing history, and exclude the sources they contain as the last coverage stage of Table~\ref{tab:selection}; the criterion uses only the FASHI detection counts, not the recovery of any parent source. A further 28 segments ($70$\,deg$^2$) are deficient at $p < 10^{-3}$; we retain them and report their exclusion as a robustness check (Section~\ref{subsec:robust}). No declination strip is deficient along its full length. The comparison region, the rectangle less the excluded segments, covers $2912$\,deg$^2$. Because its coverage is established from the detections rather than from an exposure map, what we measure below is H\,{\sc i} catalog recovery within this empirically defined comparison region: we do not assume that every parent source had an equivalent observing opportunity, and the remaining variation in depth is treated through the sensitivity proxy of Section~\ref{subsec:sens}.

Within the rectangle we select H25 sources with $0.01 < z_{\rm best} < 0.0897$, evaluated at the optical-identification position. The upper bound follows from the survey itself: the FASHI DR1 frequency range corresponds to $z < 0.0880$, and Equation~(\ref{eq:vdiff}) admits hosts up to $\Delta z = 500/c = 0.00167$ above that. The lower bound is the H25 parent criterion.

Sky coverage is only half of the coverage requirement. A radio source can be counted as an H\,{\sc i} non-detection only if the 21\,cm line it would produce falls inside the band FASHI searched, so we additionally require the expected line frequency $\nu = \nu_{\rm H\,\textsc{i}}/(1+z_{\rm best})$ to lie within the published searchable range, 1305.5--1419.5\,MHz. The detections themselves confirm that range, spanning 1305.76--1419.46\,MHz within the rectangle, and no 1\,MHz interval of it is empty of detections; the one interval with a marked deficit, 1309.5--1310.5\,MHz, with four detections against a local median of 20, lies inside the $5$\,MHz margin of the conservative-band check (Section~\ref{subsec:robust}). This criterion is applied identically to every source, irrespective of outcome, and removes 695 sources, of which one is an accepted association whose host redshift places the predicted line marginally outside the band.

A third requirement follows from the matching criterion rather than from the survey. A host with only a photometric redshift cannot satisfy Equation~(\ref{eq:vdiff}), since a typical uncertainty of $\sigma_z \sim 0.002$ corresponds to $\sim\!600$\,km\,s$^{-1}$. Such a source occupies the denominator of any recovery fraction while having almost no opportunity to enter the numerator, and the effect is directly measurable: of the 1317 photometric-redshift hosts satisfying the coverage criteria, 6 are recovered, $0.46$ per cent ($0.17$--$0.99$), against $14.54$ per cent ($14.11$--$14.98$) for spectroscopic hosts. Photometric hosts are also three times more common among radio AGN than among star-forming sources, so retaining them would bias the comparison between the classes before any gas physics entered. We therefore require a spectroscopic host redshift for the parent.

The resulting comparison parent contains 25\,573 sources, of which 1542 are classified as radio-loud AGN, 24\,031 as star-forming and none as radio-quiet quasars. Of the 3723 H25-classifiable associations with spectroscopic hosts in the band-limited rectangle, 3718 lie within it and five fall in the excluded segments. Retaining photometric hosts instead would give 26\,890 sources (1787 and 25\,103); that variant is discussed in Sections~\ref{subsec:rates} and \ref{subsec:robust}, and the alternative coverage choices --- retaining the excluded segments, excluding the marginal ones, and a more conservative band margin --- are reported as robustness checks in Section~\ref{subsec:robust}.

\subsection{FASHI Sensitivity}
\label{subsec:sens}
The probability that a radio source is recovered in H\,{\sc i} depends on the local sensitivity of the 21\,cm survey, which must therefore be controlled. FASHI DR1 provides a noise estimate only at the positions of its own detections, so we model the noise field from those 18\,045 measurements and evaluate it at the position of every parent source.

The catalog tabulates for each detection $\sigma_{\rm rms}$, the noise of its spatially integrated spectral profile in mJy at the native resolution of 6.4\,km\,s$^{-1}$ \citep{2024SCPMA..6719511Z}. The catalogued signal-to-noise ratio is recovered as $S_{\rm bf}/(\sigma_{\rm rms}\sqrt{W_{50}\,\delta v})$ with $\delta v = 6.4$\,km\,s$^{-1}$, to a median ratio of $1.0000$ with a normalised median absolute deviation of $0.0022$ over all 41\,741 detections, which confirms that $\sigma_{\rm rms}$ is a per-channel noise but does not identify the aperture it refers to. That aperture is the source's own: each detection is integrated within an ellipse about $2.0$--$2.5$ times the area of its $4.5\sigma$ region \citep{2024SCPMA..6719511Z}, so $\sigma_{\rm rms}$ grows with the size of the source. Expressing the ellipse area as a number of beams, $N_{\rm beam}$ (its area divided by the beam solid angle $1.133\,\theta^2$ for $\theta = 2.9'$, and at least one), $\log\sigma_{\rm rms}$ rises at fixed frequency with a slope of $0.654$ in $\log N_{\rm beam}$, and the aperture accounts for 49 per cent of the variance of $\log\sigma_{\rm rms}$, against 17 per cent for frequency. The catalogued values, with a median of $1.35$\,mJy for the detections in the rectangle, are therefore not a per-beam sensitivity and should not be compared directly with the median detection sensitivity of $\simeq 0.76$\,mJy\,beam$^{-1}$ quoted by \citet{2024SCPMA..6719511Z}.

Much of the apparent frequency dependence of $\sigma_{\rm rms}$ is associated with this aperture dependence. The catalogued values rise from a median of $1.08$\,mJy below 1330\,MHz to $2.31$\,mJy above 1410\,MHz, a factor of $2.1$, partly because nearby sources, observed at high frequency, are larger. After applying the empirical aperture correction $\sigma_{\rm rms}N_{\rm beam}^{-0.654}$, which we use as an approximate per-beam equivalent rather than an absolute per-beam calibration, the rise is a factor of only $1.12$, from $0.433$ to $0.487$\,mJy: under the adopted scaling, the logarithmic amplitude of the frequency trend is reduced by about 85 per cent. The per-beam values have a median of $0.46$\,mJy; their absolute level depends on the adopted aperture scaling ($0.59$\,mJy for $N_{\rm beam}^{-0.5}$), and only their relative variation enters the analysis below.

The per-beam noise is spatially coherent. After removing the frequency trend, its logarithm correlates with that of the nearest detection with a Pearson coefficient $r = 0.635$ (all correlation coefficients in this section are Pearson), against $0.019$ for a randomly shuffled control and $0.333$ for the catalogued values, at a median neighbour separation of $0.15^\circ$. We therefore model the logarithm of the per-beam noise as a median trend in 5\,MHz frequency bins plus the median residual of the eight nearest FASHI detections, and interpolate it onto each parent source. Holding out 20 per cent of the detections, the frequency trend alone reproduces $2.5$ per cent of the variance of the held-out values and the combined model $43.0$ per cent, with the root-mean-square error falling from $0.131$ to $0.101$\,dex: the field is governed mainly by sky position rather than by frequency. The interpolated per-beam noise has a median of $0.45$\,mJy and a 5th to 95th percentile range of $0.29$--$0.61$\,mJy at the parent positions, where the nearest FASHI detection lies $0.16^\circ$ away at the median and $1.20^\circ$ at worst.

Converting to the H\,{\sc i} mass detectable at signal-to-noise ratio 5 by a source unresolved by the beam, $M_{\rm lim}/M_\odot = 2.356\times10^{5}\,(1+z)^{-1}\,D_L^2 \cdot 5\sigma\sqrt{W\,\delta v}$, with $D_L$ in Mpc, $\sigma$ the per-beam noise in Jy and an assumed line width $W = 200$\,km\,s$^{-1}$, gives a median limit of $\log_{10}(M_{\rm lim}/M_\odot) = 9.10$, with a 5th to 95th percentile range of $8.35$--$9.43$; extended sources require correspondingly more flux. Varying the assumed line width between 100 and 300\,km\,s$^{-1}$ shifts the limit by $0.24$\,dex.  These per-beam noise values and mass limits are model-dependent estimates, set by the adopted aperture scaling, the interpolated noise field and the assumed line width; they serve here only to construct the relative proxy below.

Because $\log M_{\rm lim}$ is by construction $2\log D_L(z)$ plus a noise term, it correlates with redshift at $r = 0.934$, and we avoid including both quantities in one model because of that strong correlation. We therefore use a redshift-detrended version of $\log M_{\rm lim}$, $S$, obtained by subtracting its median within redshift strata of width $0.005$. Its residual linear correlation with redshift is negligible, $r = -0.008$, although median subtraction does not make it strictly independent of redshift: within each stratum it retains a small dependence on distance. $S$ has a standard deviation of $0.098$\,dex. It is an approximate sensitivity proxy, a relative measure of local depth rather than a completeness limit. It does carry information about recovery --- its median is $-0.023$\,dex among recovered sources against $0.000$\,dex among the rest. The H25 radio-AGN and star-forming populations have similar values of the proxy, with medians differing by $0.002$\,dex, and distributions that a Kolmogorov--Smirnov test does not distinguish ($p = 0.89$). This does not by itself exclude sensitivity-related selection effects that the proxy does not capture, but $S$ enters both models of Section~\ref{sec:results} as a covariate, and the fitted results are stable under the alternative noise prescriptions tested: replacing it with the same quantity built from the catalogued $\sigma_{\rm rms}$, or with an aperture scaling of $N_{\rm beam}^{-0.5}$, changes no fitted odds ratio by more than $0.03$ (Section~\ref{subsec:robust}).

Two limitations of this treatment should be stated. The noise is measured only where FASHI made a detection, so the model describes the noise field where sources were found rather than at arbitrary positions, and it represents about 43 per cent of the variance of the per-beam noise. And no sensitivity is assigned outside the verified band: the 695 sources whose expected line frequency falls outside it are excluded from the parent altogether (Section~\ref{subsec:parent}) rather than assigned an extrapolated value.

\subsection{The H25 Radio-AGN Classification}
\label{subsec:agn}
Separating accretion-powered from star-formation-powered radio emission is difficult at the luminosities probed here. Rather than construct our own radio-excess estimator, we adopt the published classification of \citet{2025MNRAS.539.1856H} (H25) as a uniform reference classification for the LoTSS DR2 parent. Its interpretation in our sample requires caution, particularly because $W3$ photometric uncertainties are substantially poorer among sources assigned to the radio-AGN class. H25 classify 963\,764 LoTSS DR2 sources with total flux density above 1.1\,mJy, an optical identification, a redshift, $z>0.01$, and WISE information, using the relation between 144\,MHz luminosity and WISE absolute magnitude calibrated against the spectroscopic classifications of \citet{2024MNRAS.534.1107D}. Sources detected in $W3$ that lie below the dividing line $\log_{10}(\Lfoursp/{\rm W\,Hz^{-1}}) = 14 - M_{W3}/2.5$ and have $\log_{10}(\Lfoursp/{\rm W\,Hz^{-1}}) < 24.8$ are rejected as star-forming; we use their inclusive radio-loud AGN sample (\texttt{AGN\_BROAD}).

Two properties of the classification limit how much weight it can carry in this regime. First, the label is a deterministic function of two quantities: evaluating $\log_{10}\Lfoursp - (14 - M_{W3}/2.5)$ reproduces \texttt{AGN\_BROAD} for $99.984$ per cent of the parent sources, the exceptions being four candidates lying just below the line. It is a diagonal cut in the $\Lfoursp$--$M_{W3}$ plane rather than a diagnostic independent of those two quantities, a point we return to in Section~\ref{subsec:surface}. Second, it depends on $W3$ photometry that is markedly poorer for the sources it selects: in the comparison parent, all 138 sources carrying no $W3$ uncertainty at all are classified as radio-loud AGN and none star-forming, and 430 of the 1542 classified AGN ($27.9$ per cent) have either no $W3$ uncertainty or a signal-to-noise ratio below five, against 30 of the 24\,031 star-forming sources ($0.12$ per cent). The median $W3$ signal-to-noise ratio is $9.0$ among the candidates and $46.9$ among the star-forming sources. The sample is therefore not immune to the bias H25 identify, whereby $W3$ non-detections default to the AGN sample. The mid-infrared colours of the H\,{\sc i}-detected hosts themselves are described in Appendix~\ref{app:wise}.

Optical spectroscopy provides an independent check, and the H25 catalog supplies one directly: it includes the classifications of \citet{2024MNRAS.534.1107D}, among them a BPT class derived from the SDSS emission lines, which we use in preference to applying demarcations of our own. Across the comparison parent, $89.1$ per cent of the H25 radio-AGN candidates carry a BPT label, of which $40.5$ per cent ($37.9$--$43.1$) fall in the Seyfert or LINER region, against $9.4$ per cent ($9.1$--$9.8$) of the star-forming sources. Across the parent population, the radio classification therefore tracks optical emission-line activity.

Among the sources FASHI recovers, that contrast largely disappears. Of the 143 recovered radio-AGN candidates, 120 carry a BPT label, and only $16.7$ per cent ($10.5$--$24.6$) fall in the Seyfert or LINER region --- against $40.5$ per cent in the parent --- while $70.8$ per cent are optically star-forming. The corresponding figure among recovered star-forming sources is $7.6$ per cent ($6.7$--$8.6$). Requiring a 21\,cm detection therefore selects, from within the H25 radio-AGN class, mostly optically star-forming hosts. We accordingly treat the recovered radio-AGN fraction as a nominal catalog-classification fraction, useful for comparison with samples classified the same way, and not as a census of accretion-powered sources. The test is one-sided: it identifies radiative-mode activity but cannot exclude the jet-mode and low-excitation radio AGN that are optically quiescent by construction, and composite objects, which lie between the two standard demarcations, are counted outside the AGN region above.

\subsection{Host Stellar Masses and Star Formation Rates}
\label{subsec:hosts}
The LoTSS value-added catalog provides stellar masses, but its reliability flag is set for only 11 per cent of the recovered parent sources against 33 per cent of the parent, which would make any host-controlled comparison unrepresentative. We therefore adopt host properties from the medium-deep GALEX--SDSS--WISE Legacy Catalog, GSWLC-M2 \citep{2018ApJ...859...11S}, which provides stellar masses and star formation rates from joint ultraviolet, optical and mid-infrared spectral-energy-distribution fitting for 361\,328 galaxies.

We match on position with a radius of $2''$. The number of matched sources is already flat at $1''$ and changes negligibly out to $5''$, and only one source has more than one counterpart. For the parent sources with host properties, the matched redshifts agree with the LoTSS values to a median of $2.5\times10^{-5}$, with none differing by more than $0.002$. Where the LoTSS mass is flagged reliable, GSWLC and LoTSS stellar masses agree with a median offset of $0.022$\,dex and a normalised median absolute deviation of $0.107$\,dex over 3239 sources.

The host-controlled analysis is restricted to sources matched to GSWLC-M2 and passing the quality requirement described below. The resulting host-property sample covers $38.2$ per cent of the radio-selected parent. Crucially for what follows, the match fractions are nearly uniform across the subsets we compare: $38.1$ per cent of the recovered sources, $36.6$ per cent of the H25 radio AGN and $38.3$ per cent of the star-forming sources, a spread of $1.7$ percentage points. GSWLC reports fitted parameters only where its SED fit carries no quality flag, which removes 159 of the 9919 matched parent sources, or $1.6$ per cent. This leaves 9760 parent sources with host properties, 564 of them classified as radio-loud AGN. Because GSWLC-M2 is built on SDSS spectroscopic targets, the spectroscopic requirement imposed on the parent costs the host-controlled comparison almost nothing: it removes four sources and no radio AGN relative to the 9764 hosts that would remain without it.

One caveat applies to the star formation rates. GSWLC-2 constrains the SED fit with the total infrared luminosity, derived from 22\,$\mu$m where available and from 12\,$\mu$m otherwise, and its mid-infrared flag is additive, so that a value of 7 denotes a 22\,$\mu$m fit with an AGN correction applied. Since the H25 classification is itself a cut involving the WISE $W3$ (12\,$\mu$m) magnitude, star formation rates derived using $W3$ are not independent of the classification. Of the 9760 parent sources with host properties, 8818 have infrared luminosities based on 22\,$\mu$m and 897 on 12\,$\mu$m, with 45 having no mid-infrared constraint; we use the 22\,$\mu$m subset as an independent check wherever the distinction could matter.

\section{Results}
\label{sec:results}
We measure H\,{\sc i} catalog recovery for the radio-selected comparison parent of Section~\ref{subsec:parent}: whether each parent source is credited with a FASHI detection (Section~\ref{subsec:xmatch}). The parent is not a volume-limited or mass-complete census, and recovery describes which radio sources a blind H\,{\sc i} survey detects, not the H\,{\sc i} content of radio galaxies as a class. We first compare the raw recovery fractions of the two nominal H25 classes (Section~\ref{subsec:rates}), then compare the classes within the host-property range they share (Section~\ref{subsec:hostcontrol}, model $\mathcal{M}_A$), and finally ask whether 144\,MHz luminosity adds information to the same model (Section~\ref{subsec:lum}, model $\mathcal{M}_B$). Unless stated otherwise, every quantity below refers to this parent and to H\,{\sc i} \emph{recovery} as defined above.

\subsection{Raw H\,{\sc i} Recovery Fractions}
\label{subsec:rates}
We first measure, without any control for host properties, how often a radio source of each class is recovered in H\,{\sc i}; Table~\ref{tab:rawrates} collects the result. Of the 1542 radio-loud AGN in the comparison parent, 143 are matched to a FASHI detection, a recovery fraction of $9.27$ per cent (95 per cent Clopper--Pearson interval $7.87$--$10.83$); of the 24\,031 star-forming sources, 3575 are matched, $14.88$ per cent ($14.43$--$15.33$). The raw odds ratio is $0.585$ ($0.491$--$0.697$).

\begin{deluxetable}{lccc}
\tablecaption{Raw H\,{\sc i} recovery fractions in the spectroscopic
comparison parent, by nominal H25 class.
\label{tab:rawrates}}
\tablehead{\colhead{Class} & \colhead{$N$} & \colhead{Recovered} &
\colhead{Fraction (per cent)}}
\startdata
Radio-loud AGN & 1542    & 143  & $9.27$  ($7.87$--$10.83$)  \\
Star-forming   & 24\,031 & 3575 & $14.88$ ($14.43$--$15.33$) \\
\hline
All            & 25\,573 & 3718 & $14.54$ ($14.11$--$14.98$) \\
\enddata
\tablecomments{Classes are the nominal H25 catalog labels
(Section~\ref{subsec:agn}); the sample is the spectroscopic comparison parent of Section~\ref{subsec:parent}. Intervals are 95 per cent Clopper--Pearson. The raw odds ratio for recovery of a radio-loud AGN relative to a star-forming source is $0.585$ ($0.491$--$0.697$). H\,{\sc i} recovery is a joint property of the source and the FASHI selection function and is not a measure of intrinsic H\,{\sc i} content (Section~\ref{subsec:sens}).}
\end{deluxetable}

The same contrast can be read as a change in composition. Radio-loud AGN make up $6.03$ per cent ($5.74$--$6.33$) of the comparison parent but only $3.85$ per cent ($3.25$--$4.52$) of its 3718 recovered sources, so requiring a 21\,cm detection lowers the classified radio-AGN fraction by a factor of $1.57$. Retaining photometric-redshift hosts in the parent would give $1.72$, a difference arising almost entirely from the denominator rather than from any change in the H\,{\sc i} data. For reference, the fraction in the unrestricted H25 catalog is $58.7$ per cent, but a comparison with that value would combine H\,{\sc i} selection with redshift, luminosity and sky-coverage effects, so it is not an appropriate control; restricting the catalog to our redshift range alone already brings it to about 7 per cent.

Section~\ref{subsec:hostcontrol} asks how much of this raw difference remains when the two classes are compared within the host-property range they share.

\subsection{H\,{\sc i} Recovery at Fixed Host Properties}
\label{subsec:hostcontrol}
Atomic gas content is known to depend strongly on stellar mass and star formation rate, so the raw difference of Section~\ref{subsec:rates} cannot be read as a property of radio activity until those are controlled. We use the 9760 parent sources with GSWLC host properties (Section~\ref{subsec:hosts}), of which 564 are classified as radio-loud AGN.

\begin{figure}
\includegraphics[width=8.50cm]{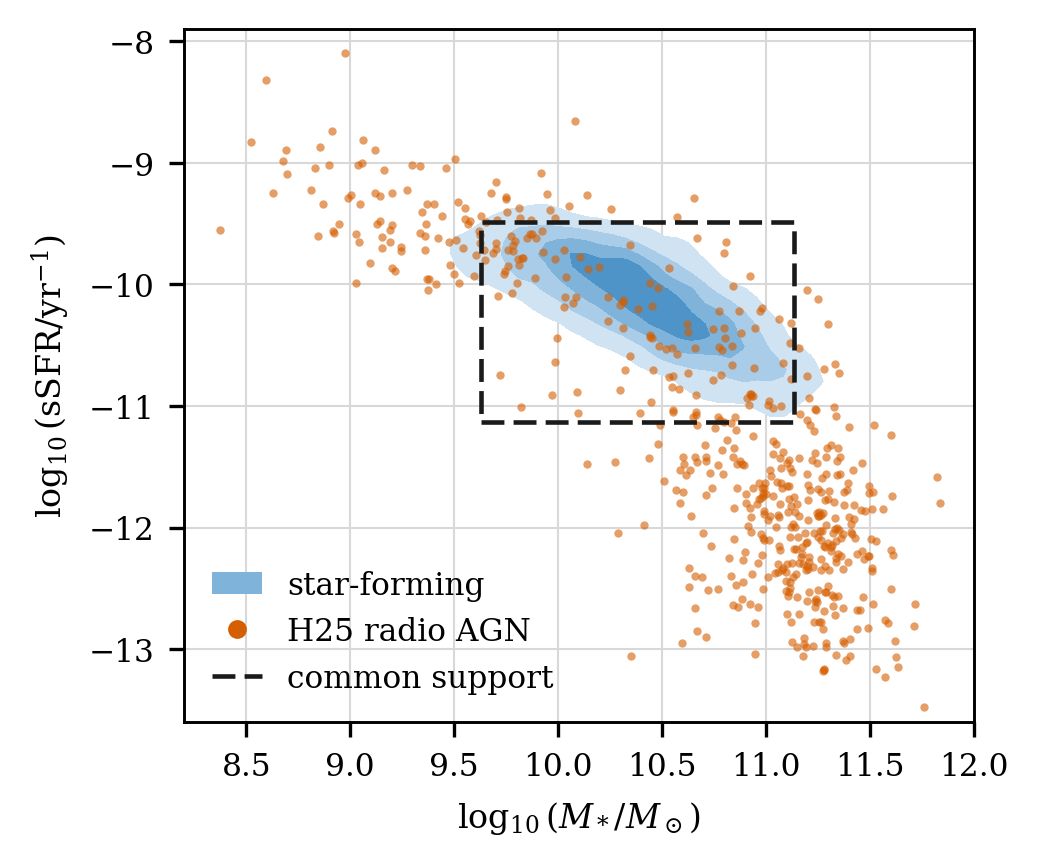}
\caption{Host stellar mass against specific star formation rate for the 9760 comparison parent sources with GSWLC host properties: the 9196 sources of the H25 star-forming class as filled contours enclosing 20, 40, 60 and 80 per cent of the distribution, and all 564 H25 radio-loud AGN as individual points. The two classes differ by $1.34$\,dex in median specific star formation rate with only $40.3$ per cent distributional overlap, estimated on 60 common bins. The dashed box marks the common support within which the host-controlled comparison of Section~\ref{subsec:hostcontrol} is possible; it contains $83.3$ per cent of the star-forming sources but only $21.6$ per cent of the radio-loud AGN. The classes are H25 radio classifications (Section~\ref{subsec:agn}), not star-forming and quiescent host classes, so radio-AGN candidates can lie at high specific star formation rate. Every candidate is drawn individually, whereas the contours omit about a fifth of the star-forming class: above that class's median specific star formation rate, 682 star-forming sources lie outside the contours, compared with 92 candidates. The 147 candidates above that median have a median stellar mass of $10^{9.6}\,M_\odot$, against $10^{10.4}\,M_\odot$ for the star-forming class; 102 of them lie within $0.3$\,dex in $\Lfoursp$ of the H25 boundary, and 58 have no $W3$ uncertainty or a $W3$ signal-to-noise ratio below five.
\label{fig:hosts}}
\end{figure}

The two classes occupy largely different regions of host-property space, as Figure~\ref{fig:hosts} shows. Their median specific star formation rates differ by $1.34$\,dex, $\log_{10}{\rm sSFR} = -11.48$ for the radio AGN against $-10.13$ for the star-forming sources, with a Kolmogorov--Smirnov statistic of $0.550$ and only $40.3$ per cent overlap between the normalised distributions. Stellar masses differ by $0.52$\,dex with $52.9$ per cent overlap. Redshifts are well matched: a median difference of $0.002$ and $86.2$ per cent overlap. A comparison at fixed host properties is therefore possible only over the range both populations occupy, $9.63 < \log_{10}(\Mstar/M_\odot) < 11.13$ and $-11.13 < \log_{10}{\rm sSFR} < -9.49$, which contains $83.3$ per cent of the star-forming sources but only $21.6$ per cent of the radio AGN. These limits are the intersection of the 5th--95th percentile ranges of the two classes, which in both quantities is the star-forming range; they were fixed before the flagged drift segments were excluded, and recomputing them on the final sample moves each by less than $0.01$\,dex.

Within that region we fit two nested logistic models to the same sources. The first, $\mathcal{M}_A$, is
\begin{equation}
\begin{split}
\mathrm{logit}\,P = & \, \alpha + \beta_{\rm class}\,\mathrm{[AGN]} + \beta_M \log_{10}\Mstar \\
	& + \beta_{\rm sSFR} \log_{10}{\rm sSFR} + \beta_z z + \beta_\sigma S,
\end{split}
\label{eq:logit}
\end{equation}
where $P$ is the probability that a parent source is recovered, $\mathrm{[AGN]}$ is a binary indicator for the H25 radio-loud AGN class and $S$ is the approximate sensitivity proxy of Section~\ref{subsec:sens}. The second, $\mathcal{M}_B$, adds $\beta_L \log_{10}\Lfoursp$ to the right-hand side of Equation~(\ref{eq:logit}) and is the subject of Section~\ref{subsec:lum}. All continuous terms are centred at their sample medians, which changes only the intercept. Coefficients are obtained by iteratively reweighted least squares, with covariances from the observed information and confidence intervals from $\exp(\beta \pm 1.96\,\sigma_\beta)$. Variance inflation factors are at most $2.2$ in $\mathcal{M}_A$ and $3.9$ in $\mathcal{M}_B$, in both cases for stellar mass, and Table~\ref{tab:robust} summarises the two models.

The key quantity in $\mathcal{M}_A$ is $\exp(\beta_{\rm class})$. The common support contains 7786 sources, 122 of them radio-loud AGN; 1112 are recovered, including 18 of the radio-loud AGN. Both models are fitted to all 7786. The class odds ratio is $1.60$ ($0.87$--$2.94$), and adding the class term to a model containing only the host and sensitivity terms gives a likelihood-ratio statistic of $2.18$ for one degree of freedom ($p = 0.14$). Without host control the odds ratio on the same sources is $1.04$ ($0.63$--$1.72$), against $0.557$ ($0.414$--$0.748$) on the full host sample. In short, nominal radio-AGN candidates have a lower raw recovery fraction, but no residual difference between the classes is statistically established within the host-property range they share.

The host-property terms of the same fit are strong. Specific star formation rate predicts recovery with an odds ratio of $4.24$ ($3.20$--$5.61$) per dex at fixed stellar mass, redshift and sensitivity, and stellar mass with $4.44$ ($3.37$--$5.86$) per dex at fixed specific star formation rate. Redshift enters as $0.43$ ($0.41$--$0.45$) per $0.01$ in $z$, and the sensitivity proxy as $0.74$ ($0.69$--$0.79$) per standard deviation of $S$ in the fitted sample, $0.097$\,dex. The two classes differ strongly in specific star formation rate (Figure~\ref{fig:hosts}), which makes host differences a plausible source of the raw contrast of Section~\ref{subsec:rates}. The model does not show that they account for all of it, since it is fitted only where the classes overlap and most radio-AGN candidates lie outside that region.

Two limits of this comparison should be stated. It covers only $21.6$ per cent of the radio AGN and does not include most quiescent radio-AGN hosts; and with 122 radio AGN in the common support, 18 of them recovered, the class odds ratio is poorly constrained, with a 95 per cent interval of $0.87$--$2.94$. The absence of a statistically significant class term therefore does not establish equivalent recovery probabilities.

\subsection{Radio Luminosity at Fixed Host Properties}
\label{subsec:lum}
Model $\mathcal{M}_B$ asks whether radio luminosity adds information to $\mathcal{M}_A$, and on the common support it does. Adding $\log_{10}\Lfoursp$ improves the fit by a likelihood-ratio statistic of $25.1$ for one degree of freedom ($p = 6\times10^{-7}$), and the odds of recovery rise by a factor of $2.03$ ($1.54$--$2.68$) per dex in $\Lfoursp$. At fixed measured stellar mass, specific star formation rate, redshift and sensitivity proxy, radio luminosity is therefore positively associated with H\,{\sc i} catalog recovery. This common-support coefficient is our primary luminosity result. Fitting the same two models to all 9760 sources with host properties extends the comparison beyond the range where the classes overlap and gives $1.77$ ($1.42$--$2.20$) per dex, with a likelihood-ratio statistic of $25.7$; we report it as an extension of the primary result, not as an interchangeable estimate.

Two features matter for how this is read. Collinearity is modest (Section~\ref{subsec:hostcontrol}), but the coefficient is an association at fixed host properties rather than a marginal effect: radio luminosity shares information with both host terms, whose odds ratios fall from $4.44$ to $2.22$ per dex for stellar mass and from $4.24$ to $2.64$ per dex for specific star formation rate when it is included. And the class coefficient moves toward unity. On the same 7786 sources, $\exp(\beta_{\rm class})$ is $1.08$ ($0.58$--$2.03$) in $\mathcal{M}_B$, against $1.60$ ($0.87$--$2.94$) in $\mathcal{M}_A$; because the H25 label is defined by a cut in the plane of $\Lfoursp$ and $M_{W3}$ (Section~\ref{subsec:agn}), the two terms carry overlapping information, and neither model establishes a residual class difference.

The luminosity coefficient describes a population dominated by star-forming sources, which make up 7664 of the 7786 sources in the common support. Fitted to those sources alone, $\mathcal{M}_B$ without its class term gives $2.21$ ($1.66$--$2.93$) per dex (Table~\ref{tab:robust}). The 122 radio-AGN candidates, with 18 recoveries, constrain a coefficient of their own only weakly, $0.16$ ($0.02$--$1.20$) per dex, and over all 564 candidates with host properties, with 50 recoveries, it is $0.29$ ($0.11$--$0.78$). The positive association therefore describes the star-forming-dominated population and is not a result about radio AGN, among which these fits show no positive association.

\begin{figure}
\includegraphics[width=8.50cm]{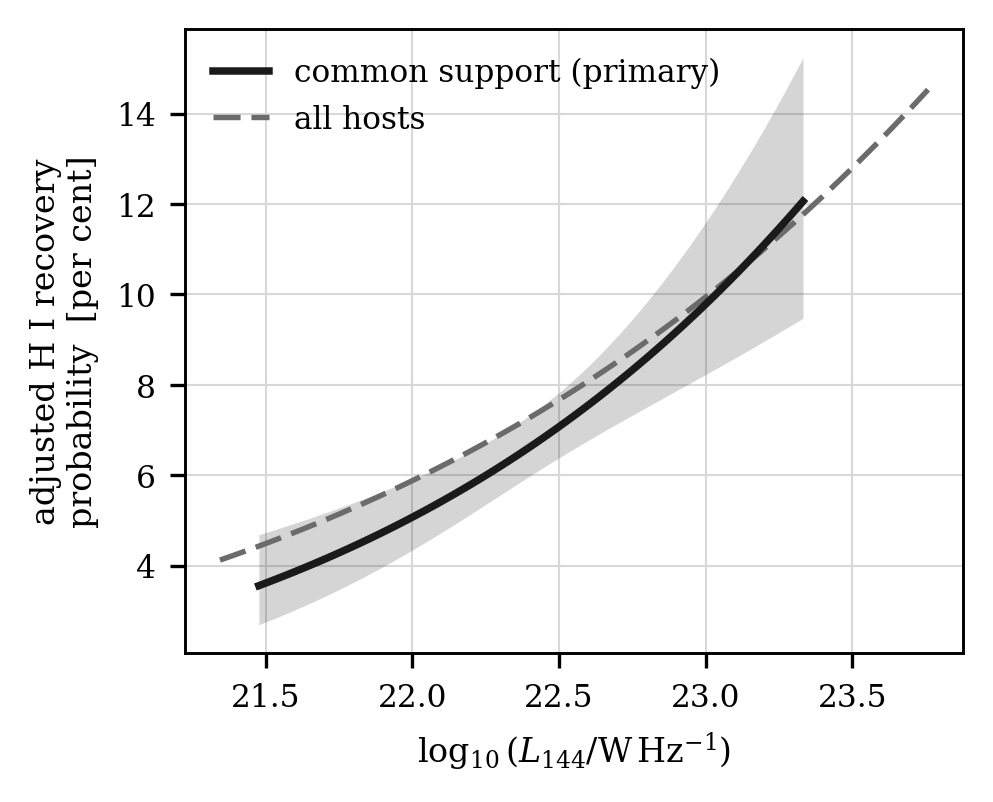}
\caption{Adjusted probability of H\,{\sc i} recovery against 144\,MHz luminosity from model $\mathcal{M}_B$, evaluated for the star-forming class at the median stellar mass, specific star formation rate, redshift and sensitivity proxy of each sample. The solid curve is the common-support fit adopted as primary; the dashed curve repeats the fit on all 9760 sources with host properties and is shown without its band. The shaded region is the 95 per cent confidence interval on the fitted probability, not a prediction interval for an individual source. Curves are drawn between the 1st and 99th percentiles of each sample's luminosity distribution.
\label{fig:predicted}}
\end{figure}

Figure~\ref{fig:predicted} shows the adjusted recovery probability from $\mathcal{M}_B$ against $\Lfoursp$, evaluated for the star-forming class at the median stellar mass, specific star formation rate, redshift and sensitivity proxy. Within the common support it rises from $3.6$ per cent at $\log_{10}(\Lfoursp/{\rm W\,Hz^{-1}}) = 21.5$ to $12.1$ per cent at $23.3$, a factor of $3.4$ between the 1st and 99th percentiles of the luminosity distribution. We report this as an association with detectable H\,{\sc i} emission. These data do not distinguish among its possible origins, which include errors in the SED-based star formation rates, the different timescales over which radio and SED indicators trace star formation, AGN contributions to $\Lfoursp$, survey selection and environment, and nothing here licenses a causal reading.

\subsection{Robustness}
\label{subsec:robust}
Table~\ref{tab:robust} repeats both models under the checks that bear on them, and further checks are described below. Across the primary fit and every confirmatory check, all on the common support, the $\Lfoursp$ coefficient of $\mathcal{M}_B$ lies between $1.98$ and $2.28$ per dex with an interval excluding unity, the lowest lower limit being $1.51$. The class coefficient of $\mathcal{M}_A$ lies between $1.47$ and $2.05$, with an interval including unity in all but one check, and the class interval of $\mathcal{M}_B$ includes unity in all of them. The exploratory companion-exclusion check is separated in the table and is not included in these ranges.

The checks address the choices of Section~\ref{sec:data}. For coverage, we retain the 89 excluded drift segments, exclude the 28 marginal segments as well, or require the expected H\,{\sc i} line to fall at least 5\,MHz inside each band edge; for the sensitivity proxy, we build it from the catalogued noise of the integrated spectrum rather than the per-beam equivalent, or adopt an aperture scaling of $N_{\rm beam}^{-0.5}$; and for the association, we credit each detection to its nearest velocity-consistent counterpart (Section~\ref{subsec:xmatch}), which adds 194 recoveries in the common support, or tighten the criterion to $60''$ and $|\Delta v| < 300$\,km\,s$^{-1}$, which retains 1075 of the 1112. Restricting to star formation rates derived from 22\,$\mu$m photometry removes the only route by which the $W3$ magnitude used in the classification could also enter the control variables. The two checks not tabulated, excluding the marginal segments and the $N_{\rm beam}^{-0.5}$ scaling, give $\mathcal{M}_A$ class and $\mathcal{M}_B$ luminosity coefficients of $1.59$ and $2.04$, and $1.59$ and $2.03$. Replacing the linear redshift term with a restricted cubic spline of four knots, a less restrictive functional form, gives $1.62$ and $2.01$, and the nonlinear terms are not significant, with a likelihood-ratio statistic of $1.9$ for two degrees of freedom.

The one check in which the class interval of $\mathcal{M}_A$ excludes unity is the 22\,$\mu$m restriction, at $2.05$ ($1.09$--$3.87$), in the direction of higher recovery for the radio-AGN candidates. That restriction is not neutral between the classes: it removes 31 of the 122 radio-AGN candidates in the common support, none of them recovered, against 5 per cent of the star-forming sources, and in $\mathcal{M}_B$ the same check gives $1.35$ ($0.70$--$2.60$). We therefore do not read it as establishing an enhancement, but it shows that the class comparison is sensitive to this choice. Environment is controlled with the companion flag of Section~\ref{subsec:association}, which is strongly associated with the class. Including it as a covariate leaves both coefficients essentially unchanged, and the flag itself is consistent with no effect, at $1.06$ ($0.88$--$1.27$). Because the spectroscopic coverage of neighbours in the external catalog is incomplete, this shows that the fit is insensitive to the available flag, not that confusion is absent. Excluding flagged sources altogether is reported as the last row and labelled exploratory: that cut removes $28.7$ per cent of the radio-AGN candidates against $18.9$ per cent of the star-forming sources, so the movement of the class coefficient to $2.10$ ($1.07$--$4.11$) reflects differential attrition rather than a controlled comparison.

These checks bound the sensitivity of the two results to the selections we can identify. They do not establish that no uncontrolled selection remains.

\begin{deluxetable*}{lcccccc}
\tabletypesize{\footnotesize}
\tablecaption{The two nested models and their robustness. In each row $\mathcal{M}_A$ and $\mathcal{M}_B$ are fitted to the same sources. The columns give the class odds ratio $\exp(\beta_{\rm class})$ in each model and the odds ratio per dex of $\Lfoursp$ in $\mathcal{M}_B$, with 95 per cent intervals, and the likelihood-ratio statistic (LR) for adding $\log_{10}\Lfoursp$ to $\mathcal{M}_A$, for one degree of freedom.
\label{tab:robust}}
\tablehead{\colhead{Sample or check} & \colhead{$N$} & \colhead{Recovered} & \colhead{$\mathcal{M}_A$} & \colhead{$\mathcal{M}_B$} & \colhead{$\mathcal{M}_B$} & \colhead{LR} \\
\colhead{} & \colhead{(AGN)} & \colhead{(AGN)} & \colhead{class} & \colhead{class} & \colhead{$\Lfoursp$, per dex} & \colhead{}}
\startdata
\cutinhead{Two nested models}
Common support (primary)    & 7786 (122) & 1112 (18) & $1.60$ ($0.87$--$2.94$) & $1.08$ ($0.58$--$2.03$) & $2.03$ ($1.54$--$2.68$) & $25.1$ \\
All hosts (extension)       & 9760 (564) & 1418 (50) & $1.21$ ($0.86$--$1.71$) & $0.79$ ($0.54$--$1.17$) & $1.77$ ($1.42$--$2.20$) & $25.7$ \\
\cutinhead{Population-specific fit on the common support}
Star-forming sources only   & 7664 (0)   & 1094 (0)  & \nodata & \nodata & $2.21$ ($1.66$--$2.93$) & $30.2$ \\
\cutinhead{Robustness checks on the common support}
Excluded segments retained  & 8114 (126) & 1113 (18) & $1.52$ ($0.83$--$2.76$) & $1.03$ ($0.55$--$1.90$) & $1.98$ ($1.51$--$2.61$) & $23.9$ \\
Conservative band           & 7310 (111) & 1107 (17) & $1.47$ ($0.79$--$2.73$) & $1.01$ ($0.53$--$1.90$) & $2.00$ ($1.52$--$2.64$) & $23.9$ \\
Integrated-spectrum noise   & 7786 (122) & 1112 (18) & $1.60$ ($0.87$--$2.92$) & $1.09$ ($0.58$--$2.02$) & $2.00$ ($1.52$--$2.64$) & $24.3$ \\
Velocity-first assignment   & 7786 (122) & 1306 (21) & $1.65$ ($0.92$--$2.94$) & $1.04$ ($0.57$--$1.89$) & $2.28$ ($1.75$--$2.98$) & $36.8$ \\
Strict association          & 7786 (122) & 1075 (18) & $1.69$ ($0.92$--$3.10$) & $1.12$ ($0.60$--$2.10$) & $2.10$ ($1.59$--$2.78$) & $27.2$ \\
SFR from 22\,$\mu$m only    & 7345 (91)  & 1071 (18) & $2.05$ ($1.09$--$3.87$) & $1.35$ ($0.70$--$2.60$) & $2.04$ ($1.54$--$2.71$) & $24.4$ \\
Environment controlled      & 7786 (122) & 1112 (18) & $1.60$ ($0.87$--$2.93$) & $1.08$ ($0.58$--$2.03$) & $2.02$ ($1.53$--$2.67$) & $24.7$ \\
\cutinhead{Exploratory check on the common support}
Companions excluded         & 6302 (87)  & 896 (16)  & $2.10$ ($1.07$--$4.11$) & $1.52$ ($0.76$--$3.04$) & $1.82$ ($1.32$--$2.50$) & $13.6$ \\
\enddata
\tablecomments{The primary fit uses the common support of Section~\ref{subsec:hostcontrol}; the extension uses all sources with host properties. The star-forming-only row omits the class term: it fits the host and sensitivity terms, with and without $\log_{10}\Lfoursp$, to the 7664 star-forming sources of the common support (Section~\ref{subsec:lum}). Adding the class term to a model with only the host and sensitivity terms gives a likelihood-ratio statistic of $2.18$ on the common support and $1.15$ for all hosts. The robustness checks are described in Section~\ref{subsec:robust}. Retaining photometric-redshift hosts adds a single source to the common support and changes no odds ratio by more than $0.01$. The exploratory check is separated because excluding companion-flagged sources removes $28.7$ per cent of the radio-AGN candidates against $18.9$ per cent of the star-forming sources, so its movement reflects differential attrition rather than a controlled comparison.}
\end{deluxetable*}
\tabletypesize{\small}

\subsection{H\,{\sc i} Recovery across the $\Lfoursp$--$M_{W3}$ Plane}
\label{subsec:surface}
\begin{figure*}[t!]
\includegraphics[angle=0,width=17.0cm]{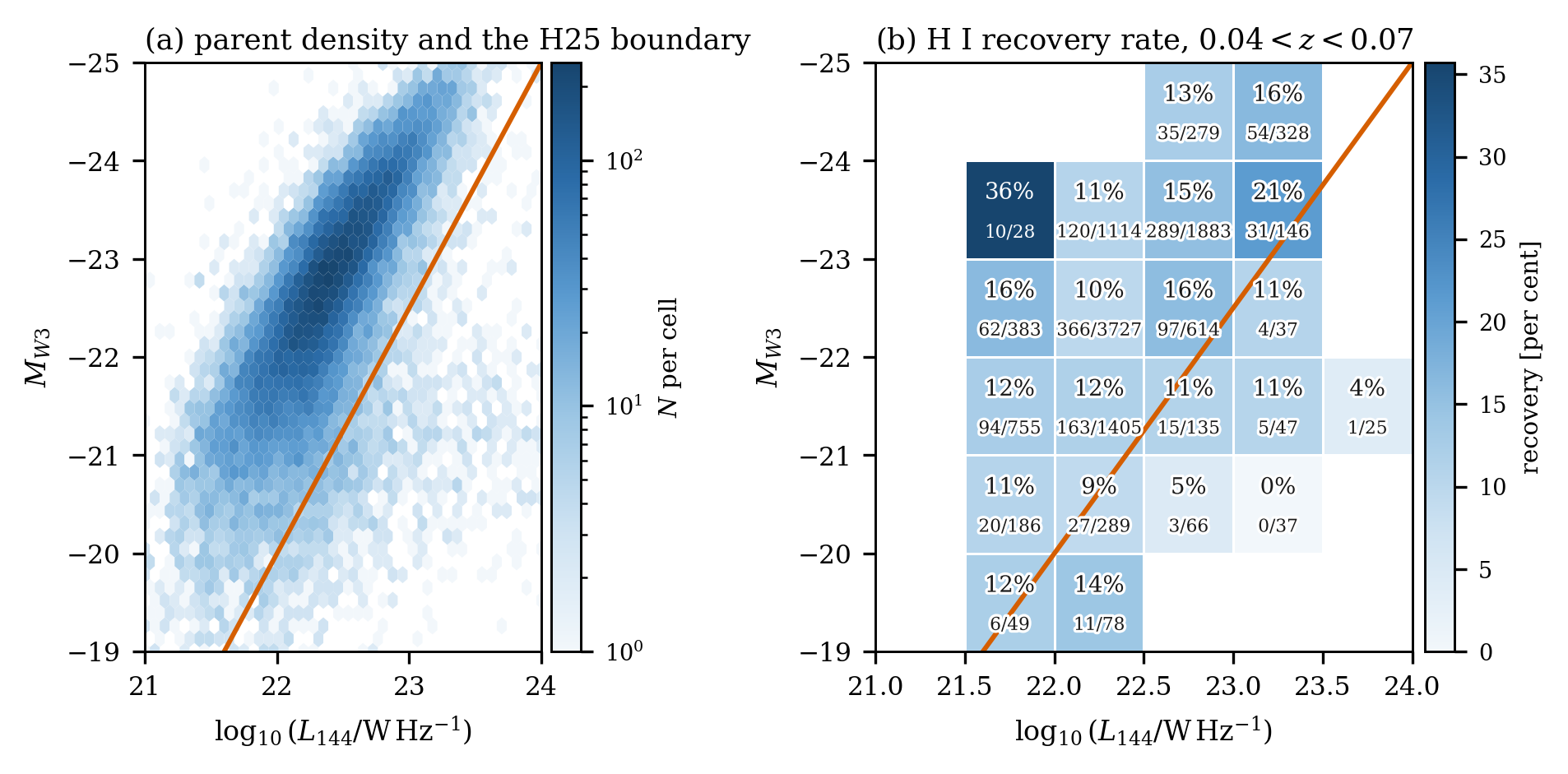}
\caption{(a) The density of the 25\,573 sources of the comparison parent in the $\Lfoursp$--$M_{W3}$ plane, with the \citet{2025MNRAS.539.1856H} star-formation exclusion boundary, $\log_{10}(\Lfoursp/{\rm W\,Hz^{-1}}) = 14 - M_{W3}/2.5$. (b) The H\,{\sc i} recovery rate on identical axes, for the 11\,849 parent sources at $0.04 < z < 0.07$, in cells containing at least 25 sources; each cell is annotated with the rate in per cent and with the number of recoveries over the number of parent sources. Twenty-one of the thirty-six cells meet that occupancy threshold. Both panels are drawn over $21.0 < \log_{10}(\Lfoursp/{\rm W\,Hz^{-1}}) < 24.0$ and $-25.0 < M_{W3} < -19.0$, with mid-infrared-fainter hosts lower down; $2.0$ and $1.3$ per cent of the sources in the two panels lie outside these limits, and blank areas within them are cells below the occupancy threshold rather than regions devoid of sources. Recovery is lowest at lower right, where radio-luminous, mid-infrared-faint hosts lie. Magnitudes are the \texttt{abs\_w3} values of the H25 catalog.
\label{fig:plane}}
\end{figure*}

Because the H25 label is a cut in the $\Lfoursp$--$M_{W3}$ plane, we close by describing how recovery varies across that plane. Figure~\ref{fig:plane} shows the binned recovery rate at $0.04 < z < 0.07$: it is lowest in the region occupied by radio-luminous, mid-infrared-faint hosts, on the radio-AGN side of the classification boundary. Over the whole comparison parent, none of the 236 sources above $10^{23.25}$\,W\,Hz$^{-1}$ with $M_{W3} > -21.5$ is recovered, a 95 per cent upper limit of $1.6$ per cent on their recovery fraction, and none of the 183 among them with a $W3$ signal-to-noise ratio of at least five is recovered either. All 236 are classified as radio-loud AGN.

We report this descriptively. The thresholds belong to a grid fixed before the rates were computed, but the region itself was identified from these data, so the result is exploratory. It is also not amenable to the host-controlled treatment of Section~\ref{subsec:hostcontrol}: only 92 of the 236 sources have host properties, with a median specific star formation rate of $10^{-12.1}$\,yr$^{-1}$, and only one of them lies within the range both classes occupy. We therefore cannot separate a radio-related effect from the quiescent host population that fills the region, nor from H\,{\sc i} masses lying below the FASHI limit.

\subsection{Association Quality and Beam Confusion}
\label{subsec:association}
The FAST beam is $\sim 2.9'$ FWHM \citep{2019SCPMA..6259502J}, corresponding to a projected diameter of $173$\,kpc at $z = 0.051$, near the middle of our redshift range, and $286$\,kpc at $z = 0.088$ in our adopted cosmology. On these scales more than one H\,{\sc i}-bearing galaxy can lie within the beam, and the measured line flux may then include emission from a companion or from a group environment. This is the dominant systematic affecting single-dish H\,{\sc i} measurements of individual galaxies \citep{2015MNRAS.449.1856J, 2016MNRAS.457.4393J}.

An assigned association does not establish that all of a detection's H\,{\sc i} emission originates in the assigned host, and neither does the strict association criterion of Section~\ref{subsec:robust}. That criterion improves positional and velocity agreement, selecting sources for which the H\,{\sc i} detection is well matched to the adopted host, but a $60''$ threshold lies inside the half-power radius of the FAST beam ($\approx 87''$), so a companion within the beam is not excluded by it. We therefore describe such samples as spatially and kinematically matched, not as unconfused. Because the FASHI centroid uncertainty depends on signal-to-noise, a separation cut also preferentially removes fainter detections, so the sources it removes are not a random subset.

Confusion is demonstrably present. Searching the full FASHI DR1 catalog rather than only our matched subset, 131 of the 4392 accepted associations ($2.98$ per cent) lie within one beam FWHM ($174''$) of another FASHI detection with $|\Delta v| < 300$\,km\,s$^{-1}$, and 95 ($2.16$ per cent) within the half-power radius. That test is nonetheless blind to the case that matters most: it finds only pairs FASHI resolved into two detections, so two galaxies whose H\,{\sc i} blends into a single detection leave no second entry to find.

We therefore repeated the search against an external spectroscopic catalog, GSWLC-M2, using the same beam radii and velocity window and excluding the host's own entry. The flag is computed for every parent source from the external catalog alone, independently of whether a FASHI counterpart exists. Among the parent sources with host properties --- the subset that lies within the footprint of the external catalog, since a GSWLC match requires an SDSS spectrum --- $44.7$ per cent of the H25 radio-AGN candidates have a companion within one beam FWHM against $20.4$ per cent of the star-forming sources, and $22.9$ against $9.1$ per cent within the half-power radius. Radio-AGN candidates are thus about twice as likely to share the beam with a spectroscopically confirmed neighbour. The absolute fractions are lower limits set by the spectroscopic completeness of the external catalog.

The external search is the more informative of the two, because it does not require the companion itself to be an H\,{\sc i} detection and so remains sensitive where companions are gas-poor.

Environment, so measured, is strongly associated with the radio-AGN class and only weakly with recovery. Among the same sources, the radio-AGN fraction rises from $4.1$ per cent for those with no spectroscopic companion within one beam FWHM to $9.1$, $18.4$ and $37.1$ per cent for those with one, two, and three or more, while the recovery fraction changes only from $14.7$ to $14.9$, $10.5$ and $10.1$ per cent. Entered as a covariate in both models on the common support, the companion flag is consistent with no effect, at $1.06$ ($0.88$--$1.27$) in $\mathcal{M}_A$, and leaves both principal results essentially unchanged (Section~\ref{subsec:robust}). The reported fits are therefore insensitive to the available flag. Because the external catalog does not record every neighbour, this does not show that confusion is absent or unimportant, and unresolved confusion would tend to raise the H\,{\sc i} flux attributed to sources in denser environments. We have not inspected individual systems, which would be the next step in bounding object-level association errors.

\section{Discussion}
\label{sec:discussion}

\subsection{What a Blind H\,{\sc i} Survey Recovers}
\label{subsec:whatselected}
Among the 3718 recovered sources of the spectroscopic comparison parent, $3.85$ per cent ($3.25$--$4.52$) carry a nominal radio-AGN classification, against $6.03$ per cent ($5.74$--$6.33$) in the parent itself, which is drawn from the same comparison region, redshift range and flux limit: a factor of $1.57$. Requiring a 21\,cm detection therefore shifts the classified composition of a low-frequency radio sample, but by less than a factor of two rather than the order of magnitude that a comparison against the unrestricted catalog would suggest. Most of that apparent order of magnitude is redshift, since the H25 catalog extends far beyond the FASHI volume.

Section~\ref{subsec:hostcontrol} shows how this contrast changes as the comparison is restricted and adjusted. The raw odds ratio of $0.585$ ($0.491$--$0.697$) is $0.557$ ($0.414$--$0.748$) on the subsample with measured host properties and $1.04$ ($0.63$--$1.72$), with no model fitted, on the sources within the range both classes occupy; with stellar mass, specific star formation rate, redshift and the sensitivity proxy included ($\mathcal{M}_A$), it is $1.60$ ($0.87$--$2.94$). The two populations are separated by $1.34$\,dex in median specific star formation rate, with only $40.3$ per cent overlap between the normalised distributions, and specific star formation rate predicts recovery at $4.24$ ($3.20$--$5.61$) per dex. Nominal radio-AGN candidates therefore have a lower raw H\,{\sc i} recovery fraction, but within the overlapping host-property range neither model establishes a statistically significant residual difference in H\,{\sc i} recovery between the nominal classes. This does not establish equivalent recovery probabilities, or that host properties account for the whole raw contrast, and the comparison does not cover most quiescent radio-AGN hosts: only $21.6$ per cent of the radio-AGN candidates lie in the shared range.

A plausible reason for the raw contrast is the construction of the label. The H25 boundary is a cut in the $\Lfoursp$--$M_{W3}$ plane, and $W3$ traces warm dust and polycyclic aromatic hydrocarbon emission, so at fixed radio luminosity a source is labelled a radio AGN more readily if its host is mid-infrared faint, and mid-infrared-faint hosts in this redshift range are predominantly quiescent. The label therefore selects, in part, the host populations in which H\,{\sc i} is least often detected. This is a property of one widely used classification as applied to this sample, not a claim that H\,{\sc i}-detected radio sources contain no accretion-powered objects.

Radio luminosity behaves differently. At fixed measured stellar mass, specific star formation rate, redshift and sensitivity proxy, it is positively associated with H\,{\sc i} catalog recovery: each dex of 144\,MHz luminosity multiplies the odds of recovery by $2.03$ ($1.54$--$2.68$) within the common support, our primary estimate, and by $1.77$ ($1.42$--$2.20$) over all sources with host properties, and the common-support coefficient lies between $1.98$ and $2.28$ across the confirmatory robustness checks. With $\Lfoursp$ in the model, the class odds ratio is $1.08$ ($0.58$--$2.03$). The association is a useful empirical result because it holds at fixed values of host properties that are themselves strongly associated with recovery: the luminosity measured by a low-frequency radio survey carries information about which of its sources a blind H\,{\sc i} survey will recover beyond what those properties provide. It describes the star-forming-dominated population of the common support rather than radio AGN (Section~\ref{subsec:lum}), and it has several possible origins: errors in the SED-based star formation rates, for which radio luminosity can act as a partial proxy; the different timescales over which radio and SED indicators trace star formation; AGN contributions to $\Lfoursp$; survey selection; and environment. Our data do not distinguish among them.

We emphasise throughout that these results concern H\,{\sc i} \emph{recovery}: whether a radio source is matched to a FASHI detection. A global 21\,cm non-detection cannot distinguish a host that is intrinsically poor in H\,{\sc i} from one whose H\,{\sc i} mass lies below the FASHI sensitivity limit at its distance, and our data do not bear on why the gas is or is not there. In particular, the luminosity association is a relationship with detectable H\,{\sc i} emission, not evidence that radio activity raises or lowers H\,{\sc i} mass, and we make no claim about gas removal, depletion or feedback.

\subsection{The H25 Classification in This Regime}
\label{subsec:agncaveat}
The H25 classification is a uniform, published labelling of a large parent catalog and is valuable for comparison with other work. Two of its properties matter for interpreting the recovery comparison (Section~\ref{subsec:agn}). It is constructed from the radio and mid-infrared observables themselves, so at fixed host properties the class term and $\Lfoursp$ carry overlapping information, as the change from $\mathcal{M}_A$ to $\mathcal{M}_B$ shows. And in this regime it depends on $W3$ photometry that is poorer for the sources it selects: $27.9$ per cent of the classified AGN have either no $W3$ uncertainty or a signal-to-noise ratio below five, against $0.12$ per cent of the star-forming sources. Since the boundary is crossed most readily by mid-infrared-faint hosts, and a non-detection is the limiting case of faintness, the classification and its principal failure mode point in the same direction.

The classification nonetheless tracks optical emission-line activity across the parent: Seyfert or LINER spectra are about four times as common among the radio-AGN candidates as among the star-forming sources (Section~\ref{subsec:agn}). That comparison concerns optical emission-line classes, and an optically star-forming spectrum, which most of the recovered radio-AGN candidates have, cannot exclude a jet-mode or low-excitation AGN. The $3.85$ per cent radio-AGN fraction among the recovered sources should therefore be read as a nominal catalog-classification fraction, reported for comparability with samples classified the same way, and not as a census of accretion-powered sources. Spectroscopic classification of the kind expected from WEAVE-LOFAR and DESI would be required to establish the radio-AGN content at these luminosities.

\subsection{The Radio-Luminous, Mid-Infrared-Faint Region}
\label{subsec:corner}
Recovery is lowest among radio-luminous, mid-infrared-faint sources, and none of the 236 parent sources above $10^{23.25}$\,W\,Hz$^{-1}$ with $M_{W3} > -21.5$ is recovered (Section~\ref{subsec:surface}). We treat this as an exploratory description, since the region was identified from these data. Almost none of its hosts lie within the range of host properties occupied by both classes, so the data cannot say whether recovery fails there because of the radio activity, because of the quiescent hosts, or because the H\,{\sc i} masses lie below the FASHI limit. Distinguishing those requires the inverse experiment: a radio-selected sample of low-redshift LoTSS sources in this part of the plane, observed in H\,{\sc i} to a known depth and analysed with censored-data methods, would convert these non-detections into upper limits and constrain the H\,{\sc i} masses directly \citep{2026ApJ...997..254Y}.

\subsection{Relation to Previous Work}
\label{subsec:prevwork}
\citet{2007A&A...464L...1E, 2010MNRAS.406..987E} found large H\,{\sc i} disks around roughly half of the hosts of nearby compact radio sources, and none around extended FR\,I sources in the same sample. Our measurement differs in every relevant respect: it records whether a blind H\,{\sc i} survey detects each member of a radio-selected parent rather than measuring H\,{\sc i} in targeted radio galaxies, it is much larger, it reaches lower radio luminosities, its structural information is catalog-level rather than morphological (Appendix~\ref{app:regression}), and 3575 of its 3718 recovered sources carry the H25 star-forming classification (Section~\ref{subsec:rates}). The absence of a structural signal in Appendix~\ref{app:regression} is therefore not in tension with their result; the two measurements are not of the same quantity, and our recovered sources are dominated by a different population from theirs. A direct comparison would require either resolved morphologies for our sources or H\,{\sc i} limits for the members of the radio-selected parent that FASHI does not recover.

\subsection{Limitations}
\label{subsec:limitations}

\begin{enumerate}
\item \textbf{Recovery, not gas content:} Every result concerns whether a radio source is \emph{recovered} by FASHI, not the intrinsic atomic gas content of its host. A global 21\,cm non-detection cannot distinguish an H\,{\sc i}-poor host from one whose H\,{\sc i} lies below the sensitivity limit at its distance.

\item \textbf{Footprint and sensitivity:} FASHI DR1 releases no coverage, exposure or noise product. The comparison region was bounded from the detection statistics of both surveys, excluding drift segments that contain no detections or significantly fewer than expected (Section~\ref{subsec:parent}), so recovery is measured within an empirically defined region rather than against a verified exposure map. The noise field was modelled from the per-beam equivalent of the catalogued noise at detection positions (Section~\ref{subsec:sens}) and represents about 43 per cent of its variance. The two classes have similar values of the resulting sensitivity proxy, whose medians differ by $0.002$\,dex, and the fitted results are stable under the alternative noise prescriptions tested; the proxy is nonetheless an approximate, model-dependent estimate rather than a measured completeness limit, and it cannot exclude sensitivity-related selection that it does not capture.

\item \textbf{Classification:} The $3.85$ per cent H25 fraction is a nominal catalog classification, constructed from $\Lfoursp$ and $M_{W3}$, whose interpretation in this sample requires caution because the $W3$ photometry is markedly poorer for the sources it assigns to the radio-AGN class (Section~\ref{subsec:agncaveat}). It is neither an upper nor a lower limit on the accretion-powered fraction. All parent sources lacking a $W3$ uncertainty are classified as AGN.

\item \textbf{Host properties:} The host-controlled analysis is restricted to the $38.2$ per cent of the parent matched to GSWLC-M2 and passing its quality requirement, with match fractions spanning only $1.7$ percentage points across the subsets compared. The comparison spans the range both classes occupy and therefore covers only $21.6$ per cent of the classified radio AGN, not most quiescent radio-AGN hosts; the absence of a statistically significant class difference within it does not establish equivalent recovery probabilities. About 9 per cent of the star formation rates use $12\,\mu$m photometry in the SED fit and are therefore not fully independent of the classification; restricting to $22\,\mu$m-based fits leaves the luminosity result unchanged but moves the class odds ratio of $\mathcal{M}_A$ to $2.05$ ($1.09$--$3.87$), through a restriction that removes a quarter of the radio-AGN candidates in the common support (Section~\ref{subsec:robust}).

\item \textbf{Association quality and beam confusion:} The FAST beam subtends $173$\,kpc at $z = 0.051$, and no separation cut excludes a companion within it, so an assigned association does not show that all of a detection's H\,{\sc i} originates in its host. Searching the full FASHI catalog flags $2.98$ per cent of associations as sharing a beam with another detection, but this counts only pairs FASHI resolved separately and remains a lower limit; the fits are insensitive to an external companion flag, but that flag is limited by spectroscopic completeness. The cross-match is one to one, with $96.1$ per cent of detections having a single candidate and a chance-association rate of $0.60$ per cent, both measured against the H25 parent rather than the full value-added catalog, so the chance rate is a lower limit and the single-candidate fraction an upper limit.
\end{enumerate}

\section{Conclusions}
\label{sec:conclusions}
We cross-matched the first FASHI data release with the LoTSS DR2 value-added catalog, defining the comparison region empirically and modelling an approximate sensitivity proxy from the detection statistics of both surveys because FASHI DR1 publishes no coverage or noise product, and measured how the H\,{\sc i} recovery of a low-redshift, radio-selected parent depends on nominal radio class, host properties and 144\,MHz luminosity. Three results follow.

\begin{itemize}
\item Nominal radio-loud AGN are recovered less often than star-forming sources in the raw comparison. Of the 1542 H25 radio AGN in the spectroscopic parent of 25\,573 sources, 143 are recovered, $9.27$ per cent ($7.87$--$10.83$), against $14.88$ per cent ($14.43$--$15.33$) of the 24\,031 star-forming sources: an odds ratio of $0.585$ ($0.491$--$0.697$). Equivalently, the nominal radio-AGN fraction falls from $6.03$ per cent ($5.74$--$6.33$) of the parent to $3.85$ per cent ($3.25$--$4.52$) of the 3718 recovered sources, a factor of $1.57$ rather than the order of magnitude obtained by comparing against the unrestricted catalog.

\item Within the overlapping host-property range, neither model establishes a statistically significant residual difference in H\,{\sc i} recovery between the nominal radio classes. The classes differ by $1.34$\,dex in median specific star formation rate, with only $40.3$ per cent overlap between the normalised distributions; on the 7786 sources of the common support, 122 of them radio AGN, the class odds ratio at fixed stellar mass, specific star formation rate, redshift and sensitivity proxy is $1.60$ ($0.87$--$2.94$), and $1.08$ ($0.58$--$2.03$) when $\Lfoursp$ is added. This does not establish equivalent recovery probabilities, and the comparison does not cover most quiescent radio-AGN hosts.

\item At fixed measured stellar mass, specific star formation rate, redshift and sensitivity proxy, 144\,MHz luminosity is positively associated with H\,{\sc i} catalog recovery. Within the common support each dex of luminosity raises the odds of recovery by a factor of $2.03$ ($1.54$--$2.68$), our primary estimate, and the coefficient remains between $1.98$ and $2.28$ across the confirmatory robustness checks; over all sources with host properties it is $1.77$ ($1.42$--$2.20$). The association is carried primarily by the star-forming population, which alone gives $2.21$ ($1.66$--$2.93$) per dex, and is not established among the radio-AGN candidates; we draw no causal inference from it.
\end{itemize}

These results describe H\,{\sc i} recovery within an empirically defined comparison region, with an approximate sensitivity proxy, rather than intrinsic gas content, and we make no claim about gas removal, depletion or feedback. The natural continuation inverts the selection: a radio-selected sample in the radio-luminous, mid-infrared-faint region where none of 236 parent sources is recovered, observed in H\,{\sc i} to a known depth and analysed with censored-data methods, would convert these non-detections into constraints on H\,{\sc i} mass \citep{2026ApJ...997..254Y}. The FASHI/LoTSS association catalog is released in machine-readable form as a first step towards the planned FAST Radio Galaxy catalog. The second FASHI data release, which covers about 19\,500\,deg$^2$ with 156\,411 detections \citep{2026arXiv260631539Z}, will allow this recovery measurement to be repeated over a larger area.

\section*{Acknowledgments}
This work is supported by the National SKA Program of China No. 2025SKA0150103, National Natural Science Foundation of China under Nos. 12550002, 12133008, 12221003, 11890692. S.K.B. and T.F. acknowledge the science research grants from the China Manned Space Project with No. CMS-CSST-2021-A04 and No. CMS-CSST-2025-A10.

This work uses data from the FASHI project. FASHI made use of data from FAST (Five-hundred-meter Aperture Spherical radio Telescope), a Chinese national mega-science facility operated by the National Astronomical Observatories, Chinese Academy of Sciences.

LOFAR data products were provided by the LOFAR Surveys Key Science project. LOFAR is the Low Frequency Array designed and constructed by ASTRON. It has observing, data processing, and data storage facilities in several countries, which are owned by various parties (each with their own funding sources), and which are collectively operated by the LOFAR ERIC under a joint scientific policy. The LOFAR resources have benefited from the following recent major funding sources: CNRS-INSU, Observatoire de Paris and Université d'Orléans, France; BMFTR, MKW-NRW, MPG, Germany; Science Foundation Ireland (SFI), Department of Business, Enterprise and Innovation (DBEI), Ireland; NWO, The Netherlands; The Science and Technology Facilities Council, UK; Ministry of Science and Higher Education, Poland; The Istituto Nazionale di Astrofisica (INAF), Italy. 

Based on observations made with the NASA Galaxy Evolution Explorer. GALEX is operated for NASA by the California Institute of Technology under NASA contract NAS5-98034.

This publication makes use of data products from the Wide-field Infrared Survey Explorer, which is a joint project of the University of California, Los Angeles, and the Jet Propulsion Laboratory/California Institute of Technology, funded by the National Aeronautics and Space Administration.

Funding for the Sloan Digital Sky Survey IV has been provided by the Alfred P. Sloan Foundation, the U.S. Department of Energy Office of Science, and the Participating Institutions.
SDSS-IV acknowledges support and resources from the Center for High Performance Computing  at the University of Utah. The SDSS website is www.sdss4.org.
SDSS-IV is managed by the Astrophysical Research Consortium for the Participating Institutions of the SDSS Collaboration including the Brazilian Participation Group, the Carnegie Institution for Science, Carnegie Mellon University, Center for Astrophysics | Harvard \& Smithsonian, the Chilean Participation Group, the French Participation Group, Instituto de Astrof\'isica de Canarias, The Johns Hopkins University, Kavli Institute for the Physics and Mathematics of the Universe (IPMU) / University of Tokyo, the Korean Participation Group, Lawrence Berkeley National Laboratory, Leibniz Institut f\"ur Astrophysik Potsdam (AIP),  Max-Planck-Institut f\"ur Astronomie (MPIA Heidelberg), Max-Planck-Institut f\"ur Astrophysik (MPA Garching), Max-Planck-Institut f\"ur Extraterrestrische Physik (MPE), National Astronomical Observatories of China, New Mexico State University, New York University, University of Notre Dame, Observat\'ario Nacional / MCTI, The Ohio State University, Pennsylvania State University, Shanghai Astronomical Observatory, United Kingdom Participation Group, Universidad Nacional Aut\'onoma de M\'exico, University of Arizona, University of Colorado Boulder, University of Oxford, University of Portsmouth, University of Utah, University of Virginia, University of Washington, University of Wisconsin, Vanderbilt University, and Yale University.

The Legacy Surveys consist of three individual and complementary projects: the Dark Energy Camera Legacy Survey (DECaLS; Proposal ID \#2014B-0404; PIs: David Schlegel and Arjun Dey), the Beijing-Arizona Sky Survey (BASS; NOAO Prop. ID \#2015A-0801; PIs: Zhou Xu and Xiaohui Fan), and the Mayall z-band Legacy Survey (MzLS; Prop. ID \#2016A-0453; PI: Arjun Dey). DECaLS, BASS and MzLS together include data obtained, respectively, at the Blanco telescope, Cerro Tololo Inter-American Observatory, NSF’s NOIRLab; the Bok telescope, Steward Observatory, University of Arizona; and the Mayall telescope, Kitt Peak National Observatory, NOIRLab. Pipeline processing and analyses of the data were supported by NOIRLab and the Lawrence Berkeley National Laboratory (LBNL). The Legacy Surveys project is honored to be permitted to conduct astronomical research on I’oligam Du’ag (Kitt Peak), a mountain with particular significance to the Tohono O’odham Nation.

NOIRLab is operated by the Association of Universities for Research in Astronomy (AURA) under a cooperative agreement with the National Science Foundation. LBNL is managed by the Regents of the University of California under contract to the U.S. Department of Energy.

This project used data obtained with the Dark Energy Camera (DECam), which was constructed by the Dark Energy Survey (DES) collaboration. Funding for the DES Projects has been provided by the U.S. Department of Energy, the U.S. National Science Foundation, the Ministry of Science and Education of Spain, the Science and Technology Facilities Council of the United Kingdom, the Higher Education Funding Council for England, the National Center for Supercomputing Applications at the University of Illinois at Urbana-Champaign, the Kavli Institute of Cosmological Physics at the University of Chicago, Center for Cosmology and Astro-Particle Physics at the Ohio State University, the Mitchell Institute for Fundamental Physics and Astronomy at Texas A\&M University, Financiadora de Estudos e Projetos, Fundacao Carlos Chagas Filho de Amparo, Financiadora de Estudos e Projetos, Fundacao Carlos Chagas Filho de Amparo a Pesquisa do Estado do Rio de Janeiro, Conselho Nacional de Desenvolvimento Cientifico e Tecnologico and the Ministerio da Ciencia, Tecnologia e Inovacao, the Deutsche Forschungsgemeinschaft and the Collaborating Institutions in the Dark Energy Survey. The Collaborating Institutions are Argonne National Laboratory, the University of California at Santa Cruz, the University of Cambridge, Centro de Investigaciones Energeticas, Medioambientales y Tecnologicas-Madrid, the University of Chicago, University College London, the DES-Brazil Consortium, the University of Edinburgh, the Eidgenossische Technische Hochschule (ETH) Zurich, Fermi National Accelerator Laboratory, the University of Illinois at Urbana-Champaign, the Institut de Ciencies de l’Espai (IEEC/CSIC), the Institut de Fisica d’Altes Energies, Lawrence Berkeley National Laboratory, the Ludwig Maximilians Universitat Munchen and the associated Excellence Cluster Universe, the University of Michigan, NSF’s NOIRLab, the University of Nottingham, the Ohio State University, the University of Pennsylvania, the University of Portsmouth, SLAC National Accelerator Laboratory, Stanford University, the University of Sussex, and Texas A\&M University.

BASS is a key project of the Telescope Access Program (TAP), which has been funded by the National Astronomical Observatories of China, the Chinese Academy of Sciences (the Strategic Priority Research Program “The Emergence of Cosmological Structures” Grant \# XDB09000000), and the Special Fund for Astronomy from the Ministry of Finance. The BASS is also supported by the External Cooperation Program of Chinese Academy of Sciences (Grant \# 114A11KYSB20160057), and Chinese National Natural Science Foundation (Grant \# 12120101003, \# 11433005).

The Legacy Survey team makes use of data products from the Near-Earth Object Wide-field Infrared Survey Explorer (NEOWISE), which is a project of the Jet Propulsion Laboratory/California Institute of Technology. NEOWISE is funded by the National Aeronautics and Space Administration.

The Legacy Surveys imaging of the DESI footprint is supported by the Director, Office of Science, Office of High Energy Physics of the U.S. Department of Energy under Contract No. DE-AC02-05CH11231, by the National Energy Research Scientific Computing Center, a DOE Office of Science User Facility under the same contract; and by the U.S. National Science Foundation, Division of Astronomical Sciences under Contract No. AST-0950945 to NOAO.

\appendix
\section{The LoTSS Component Code and Atomic Gas Content}
\label{app:regression}
This appendix collects the analysis of the LoTSS component code, which we report as a secondary result. \texttt{S\_Code} is an extraction and component-association code rather than a physical morphology classifier, so the comparison below should not be read as a test of compact against extended radio structure.

\subsection{The component code and the mass-restricted sample}
The \texttt{S\_Code} column records how a source was extracted and assembled, not its physical morphology. Following the catalog definitions, \texttt{S} denotes an isolated source fitted with a single Gaussian, \texttt{C} a source fitted with a single Gaussian within an island containing other sources, \texttt{M} a source requiring multiple Gaussians, and \texttt{Z} a composite source assembled from several radio components, largely through the visual inspection workflow described by \citet{2023A&A...678A.151H}.

We therefore refer to \emph{single-component} (S) and \emph{multi-component} (M) sources. No composite (\texttt{Z}) source satisfies the stellar-mass requirement, so the multi-component class in the mass-analysis sample consists entirely of \texttt{M}-coded sources. We do not equate these classes with jet--lobe systems, Fanaroff--Riley types, or bent or disturbed morphologies. Two consequences follow and are important for interpreting our results. First, assignment to the M class requires that a source be resolved into distinguishable components, which depends on angular size and hence on distance. Second, the component-association workflow in DR2 was applied only above a flux-density threshold, so the M class is also correlated with brightness. Both effects are visible in our sample (Sections~\ref{subsec:zdist} and \ref{subsec:ldist}). The matched sample contains no sources with \texttt{S\_Code} $=$ \texttt{C}.

The stellar-mass requirement is not structure-independent. Of the 3258 single-component sources, 490 (15.0\%) have a reliable value; of the 873 multi-component sources, 39 (4.5\%); and of the 261 composite sources, none. Composite sources are assembled from several radio components and their host identifications are correspondingly less secure, so the value-added catalog does not report a reliable stellar mass for them. This selection therefore acts directly on the component class we go on to test, and removes the composite-coded sources entirely. For that reason the stellar-mass requirement is imposed only on the sample used for the gas-fraction analysis, and not on the samples used for the radio-source composition (Section~\ref{subsec:agn}).

For the mass-analysis sample we use the atomic gas fraction $\fHI \equiv \MHI/\Mstar$, quoted as $\log_{10}\fHI$. Absolute H\,{\sc i} mass correlates strongly with galaxy mass, so the gas fraction is the more informative quantity when comparing galaxies of different stellar mass \citep{2018MNRAS.476..875C, 2022ARA&A..60..319S}. We note that H\,{\sc i} masses are taken directly from the FASHI catalog, whose tabulated distances correspond to a slightly different distance scale from the $H_0 = 70$\,km\,s$^{-1}$\,Mpc$^{-1}$ adopted for the LoTSS quantities, with flow corrections applied at the lowest velocities. The resulting systematic in $\log \MHI$ is $\sim\!0.03$\,dex, well below the scatter in any relation considered here, and affects both component classes equally.

The analysis in this appendix uses the subset of accepted associations lying within the comparison rectangle and satisfying the redshift and band criteria of Section~\ref{subsec:parent}, for which the value-added catalog reports a reliable host stellar mass (\texttt{flag\_mass}\,$=1$): 500 sources, of which 462 are \texttt{S}-coded and 38 \texttt{M}-coded, with no composite source meeting the mass requirement. We do not impose the spectroscopic-host requirement or the exclusion of the flagged drift segments here. Those requirements exist in Section~\ref{subsec:parent} to make H\,{\sc i} recovery a well-defined quantity, and every source in this sample is by construction an H\,{\sc i} detection; six of the 500 have photometric host redshifts, and excluding them changes the component-class coefficient by $0.001$\,dex.

As a robustness check we tightened the association criteria of Section~\ref{subsec:xmatch}, requiring a separation below $60''$ and $\Delta v < 300$\,km\,s$^{-1}$ between the FASHI H\,{\sc i} centroid and the optical host. This removes 27 of the 500 sources (5.4 \%), leaving 473 (437 S, 36 M). The two criteria act on different populations: the 23 sources removed by the separation criterion all have spectroscopic host redshifts and are drawn from both component classes (21 S, 2 M), consistent with the correlation between FASHI centroid offset and signal-to-noise ratio, whereas all 4 sources removed by the velocity criterion have photometric host redshifts. We retain the 500 as our primary sample, since the $60''$ and $300$\,km\,s$^{-1}$ thresholds are not externally calibrated, and quote the corresponding gas-fraction offset for the 473 among the diagnostics below.

\subsection{Gas Fraction and Stellar Mass}
\label{subsec:gmass}
\begin{figure*}[t!]
\centering
\includegraphics[angle=0,width=12.0cm]{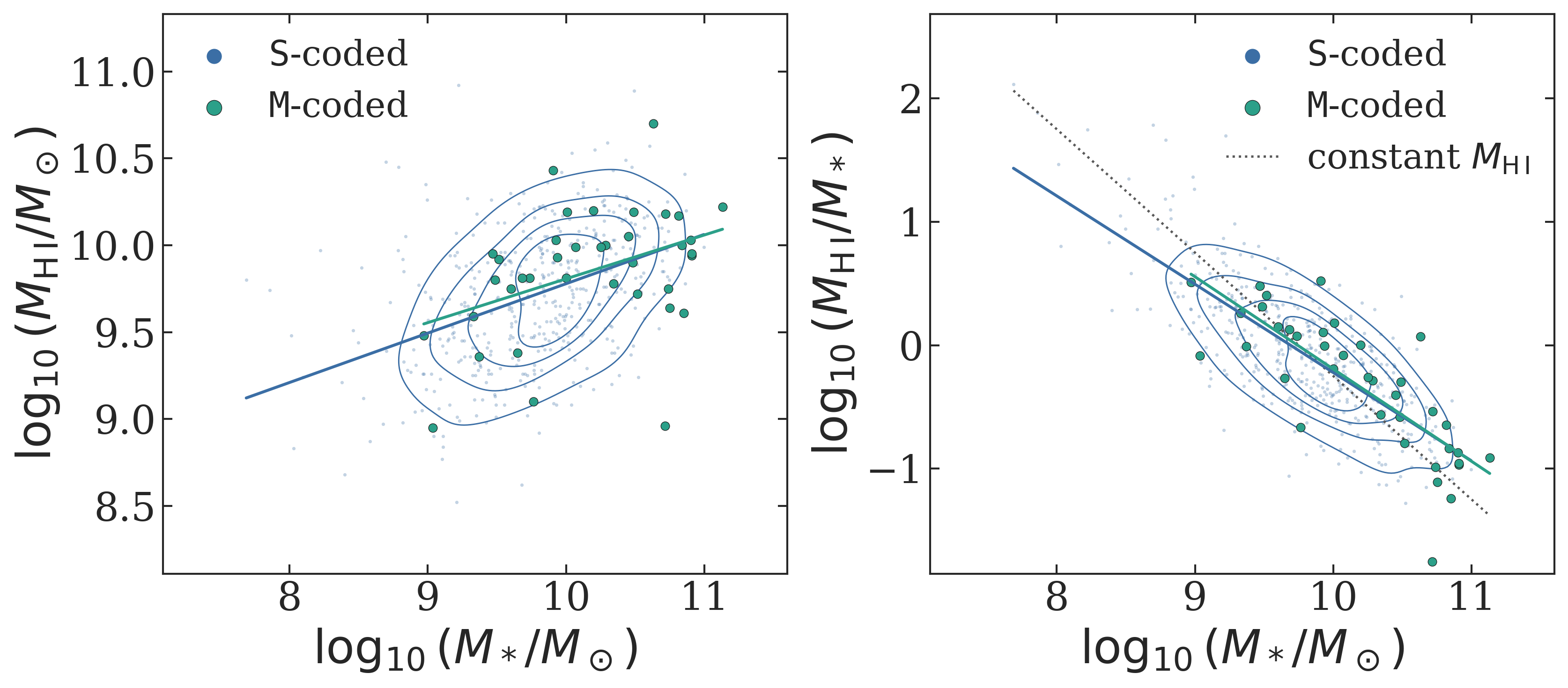}
\caption{Atomic hydrogen against host stellar mass for the 500 associations within the comparison rectangle with a reliable stellar-mass estimate: 462 \texttt{S}-coded (blue, with contours enclosing 20, 40, 60 and 80 \% of the distribution) and 38 \texttt{M}-coded (green) sources. No composite (\texttt{Z}) source satisfies the stellar-mass requirement. Left: $\log_{10}\MHI$, the response variable adopted in Section~\ref{subsec:regression}, whose fitted slopes against stellar mass are $+0.284$ (\texttt{S}) and $+0.252$ (\texttt{M}). Right: the gas-fraction plane, in which both axes share $\log_{10}\Mstar$ by construction and the slopes are correspondingly $-0.716$ and $-0.748$; the dotted line is a locus of constant $\MHI$ at the sample median. The \texttt{M}-coded sources preferentially occupy the high-mass end and show no systematic offset at fixed stellar mass in either projection.}
\label{fig:gfsmass}
\end{figure*}

Figure~\ref{fig:gfsmass} shows $\log_{10}\MHI$ and $\log_{10}\fHI$ against $\log_{10}(\Mstar/M_\odot)$ for the 462 \texttt{S}-coded and 38 \texttt{M}-coded sources. Gas fraction decreases with stellar mass across the sample (Pearson $r = -0.764$), and the two classes follow the relation with indistinguishable slopes, $-0.716 \pm 0.029$ for S and $-0.748 \pm 0.096$ for M (difference $0.032 \pm 0.100$).

Part of this anticorrelation is arithmetic rather than physical: $\log\fHI = \log\MHI - \log\Mstar$, so the two axes share a variable by construction. Fitting $\log\MHI$ against $\log\Mstar$ directly in the same sample gives a slope of $0.284$, and we adopt $\log\MHI$ as the response variable in the regression of Section~\ref{subsec:regression} for this reason. The relevant result here is not the slope but the absence of an offset between the two component classes: at fixed stellar mass the \texttt{M}-coded points lie within the \texttt{S}-coded distribution, with no systematic displacement.

The M sources do not sample the full stellar mass range. Their median stellar mass is $\log_{10}(\Mstar/\Msun) = 10.22$, and only 6 of 38 lie below 9.5, whereas 138 of 462 S sources do; the two mass distributions differ significantly (Kolmogorov--Smirnov $p = 0.001$). The M sources therefore sample preferentially the high-mass end, where gas fractions are lower in general, a consequence of the mass dependence of gas content in the local galaxy population \citep{2018MNRAS.476..875C, 2010ApJ...721..193P}. Any uncorrected comparison of $\MHI$ or $\fHI$ between the classes will therefore inherit this mass difference, which is the first reason a multivariate treatment is required.

Sixteen sources in the 500 have $\log_{10}\fHI > 1$. All are single-component, their median stellar mass is $\log_{10}(\Mstar/\Msun) = 8.74$ against 9.84 for the sample, and their FASHI centroid offsets and signal-to-noise ratios are unexceptional. Their properties are consistent with gas-rich low-mass hosts, although stellar-mass errors or association ambiguity may also contribute; normal centroid offsets and signal-to-noise ratios do not by themselves exclude a wrong host identification. We retain them in the primary analysis and quantify their influence in Section~\ref{subsec:regression}.

\subsection{Redshift Distributions}
\label{subsec:zdist}
We compare the redshift distributions of the two classes. In the accepted association sample the two classes differ significantly in redshift: the median is $z = 0.0385$ $[0.0376, 0.0393]$ for S sources and $0.0305$ [0.0297, 0.0313] for \texttt{M}- and \texttt{Z}-coded sources combined (\texttt{M}$+$\texttt{Z}; Kolmogorov--Smirnov $p = 7.5\times10^{-36}$; Mann--Whitney $p = 1.9\times10^{-46}$; 95 \% bootstrap intervals on the medians). In the 500-source mass-analysis sample the distributions are statistically indistinguishable, with medians $0.0498$ $[0.0459, 0.0529]$ and $0.0485$ $[0.0401, 0.0579]$ and $p = 0.69$ (Mann--Whitney $p = 0.88$).

In the accepted association sample this offset is expected from the way the component classes are constructed (Section~\ref{app:regression}). Resolving a source into separate components requires sufficient angular size, so nearer sources are more readily assigned to M, while more distant sources of the same intrinsic structure remain single-component. Some fraction of our S sources are therefore \texttt{M}-coded sources that are unresolved at their distance \citep{2019MNRAS.488.2701M, 2023A&A...678A.151H}. This is a limitation of the classification, not a property of the galaxies, and it means the S class is heterogeneous. The effect is strongest for the composite (Z) sources, which have the lowest median redshift ($z = 0.0245$), the largest median angular size ($78.5''$), and are resolved without exception --- and which are entirely removed by the stellar-mass requirement. Restricted to \texttt{S}- and \texttt{M}-coded sources, the difference remains significant ($p = 8\times10^{-20}$), but it does not survive into the analysis sample.

Redshift also enters through the H\,{\sc i} selection. In a flux-limited 21\,cm survey the minimum detectable $\MHI$ rises steeply with distance, so more distant detections are systematically more H\,{\sc i}-massive \citep{2011AJ....142..170H, 2018ApJ...861...49H}. Redshift in our regression is thus best understood as a proxy for the survey selection function rather than as a physical variable.

Because the two classes are indistinguishable in redshift in the sample we analyze, redshift cannot account for any difference in gas fraction between them there; it is retained in the regression of Section~\ref{subsec:regression} because it tracks the H\,{\sc i} flux limit, not because it separates the classes. Radio luminosity behaves differently: the classes differ in $\Lfoursp$ in both samples (Section~\ref{subsec:ldist}), so it is the control most likely to account for any raw difference in gas fraction.

\subsection{Radio Luminosity Distributions}
\label{subsec:ldist}
\begin{figure*}[t!]
\centering
\includegraphics[angle=0,width=12.0cm]{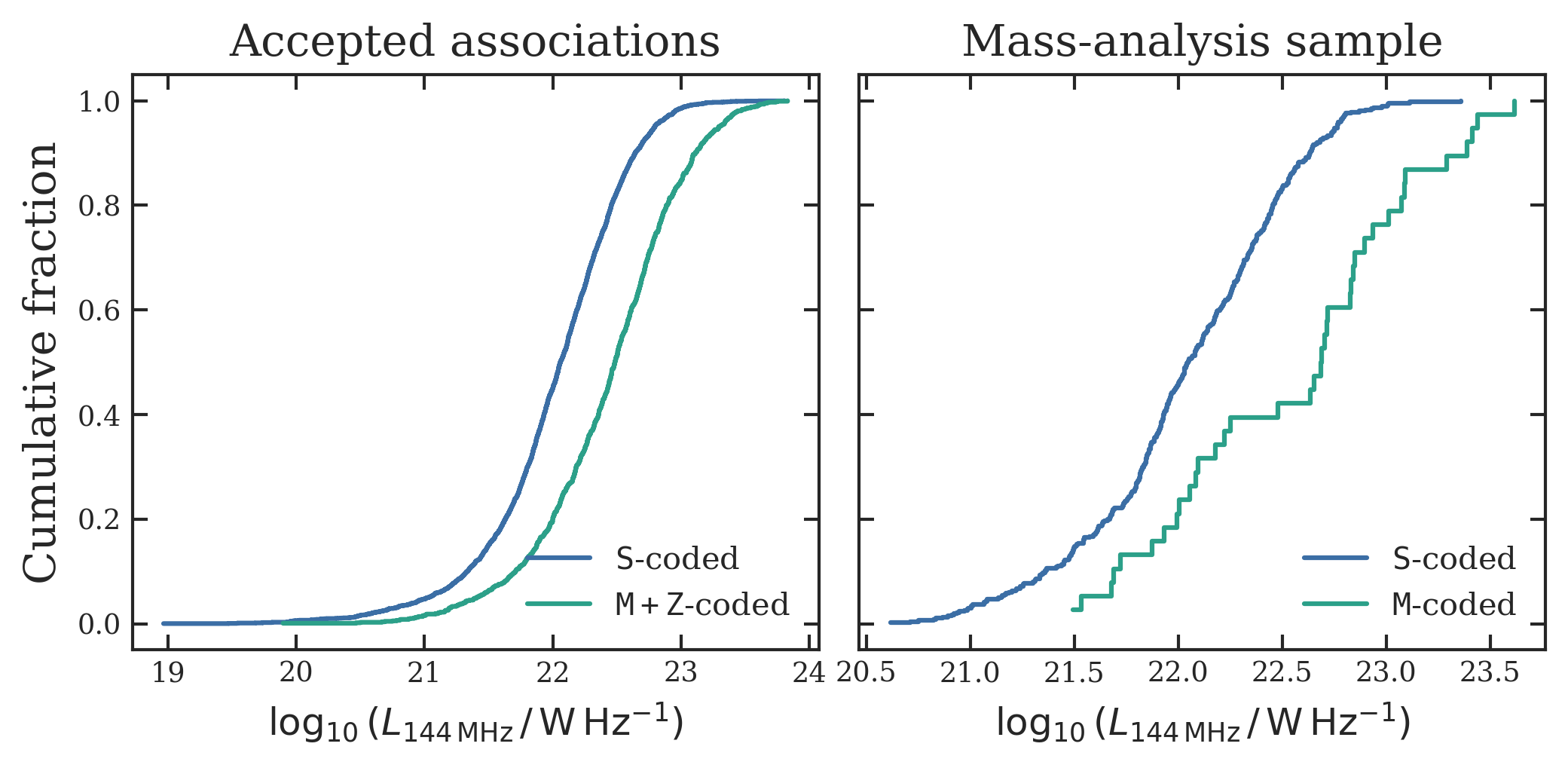}
\caption{Empirical cumulative distributions of 144\,MHz luminosity for \texttt{S}-coded (blue) and \texttt{M}$+$\texttt{Z}-coded (green) LoTSS sources. Left: the accepted association sample within the comparison rectangle (4208 sources), median offset $0.425$\,dex. Right: the 500-source mass-analysis sample, which contains no \texttt{Z} source, median offset $0.643$\,dex. The \texttt{M} class is effectively truncated below $\log_{10}\Lfoursp \approx 21.5$, reflecting the flux dependence of the component-association procedure.}
\label{fig:ldist}
\end{figure*}

Figure~\ref{fig:ldist} shows the 144\,MHz luminosity distributions. In the accepted association sample the combined \texttt{M}$+$\texttt{Z} sources have a median $\Lfoursp$ higher than that of the \texttt{S}-coded sources by $0.421$\,dex, a factor of $2.6$ ($0.425$\,dex for the 4208 sources within the comparison rectangle shown in Figure~\ref{fig:ldist}); restricted to \texttt{M}-coded sources alone the offset is $0.452$\,dex. Figure~\ref{fig:ldist} plots the \texttt{M}$+$\texttt{Z} combination in the left panel and \texttt{M}-coded sources alone in the right, where no \texttt{Z} source survives. The offset is larger in the 500-source mass-analysis sample, $0.643$\,dex or a factor of $4.4$ (medians $\log_{10}\Lfoursp = 22.046$ $[21.998, 22.120]$ and $22.689$ $[22.200, 22.836]$; Kolmogorov--Smirnov $p = 6.4\times10^{-8}$, Mann--Whitney $p = 8.5\times10^{-7}$).

The difference is not confined to the median. The M class is effectively absent at low luminosity: 67 of the 462 S sources (14.5 \%) lie below $\log_{10}\Lfoursp = 21.5$, against 1 of 38 M sources, and the faintest M source has a total flux density of $0.96$\,mJy compared with $0.37$\,mJy for S. This is at least in part the flux dependence of the component-association procedure (Section~\ref{app:regression}): assigning multi-component structure requires sufficient signal to detect the components individually, so brighter sources are more readily classified as M. The remainder may reflect a genuine difference between the physical populations. We cannot separate the two contributions here, but both point the same way --- $\Lfoursp$ is correlated with component class, by construction as well as possibly physically, and unlike redshift this correlation persists into the sample we analyze. It must therefore be controlled before any structural comparison of gas content is meaningful.

The luminosity range occupied by the sample is itself informative. Between the 5th and 95th percentiles the sample spans $\log_{10}(\Lfoursp/{\rm W\,Hz^{-1}}) = 21.16$ to $22.83$, with a median of $22.08$; only 14 of the 500 sources exceed $10^{23}$\,W\,Hz$^{-1}$ and none exceeds $10^{24}$\,W\,Hz$^{-1}$. At 144\,MHz this is the regime in which star formation and low-luminosity accretion activity both contribute substantially to the radio output \citep{2018MNRAS.475.3010G, 2021A&A...648A...6S, 2025MNRAS.539.1856H}. In the main text the two are separated, nominally, by the H25 classification (Section~\ref{subsec:agn}).

\subsection{Component Class and Atomic Gas Content}
\label{subsec:regression}
As a secondary test we ask whether the LoTSS component class carries information about the atomic gas content of the hosts, using the 500-source mass-analysis sample. The raw comparison is confounded and we report it only for completeness: \texttt{M}-coded sources have a median gas fraction $0.167$\,dex lower than \texttt{S}-coded sources (Appendix~\ref{app:regression}), an offset in the opposite sense to the expectation that morphologically complex radio sources are more gas-rich, but they also exceed them by $0.40$\,dex in median stellar mass and $0.64$\,dex in 144\,MHz luminosity. We therefore fit
\begin{equation}
\log_{10}\MHI = a + b_1\,\mathrm{[M]} + b_2 \log_{10}\Mstar + b_3 z + b_4 \log_{10}\Lfoursp + \epsilon,
\label{eq:ols}
\end{equation}
where $\mathrm{[M]}$ is a binary indicator for \texttt{M}-coded sources and $z$ is the host redshift, by ordinary least squares with heteroscedasticity-robust (HC3) standard errors.

We adopt $\log_{10}\MHI$ rather than $\log_{10}\fHI$ as the response because the latter shares $\log_{10}\Mstar$ with a predictor, so that stellar-mass measurement error enters the two sides of the equation with opposite signs. The two forms are algebraically equivalent: the coefficients on component class, redshift and luminosity are identical to one part in $10^{13}$, and only the stellar-mass coefficient differs, by exactly unity.

The fitted coefficients are $b_1 = +0.045 \pm 0.051$, $b_2 = +0.042 \pm 0.029$, $b_3 = 12.17 \pm 1.01$ and $b_4 = 0.111 \pm 0.037$, with an adjusted $R^2$ of $0.499$. The 95 \% confidence interval on the component-class coefficient is $[-0.054, +0.144]$\,dex, and a bootstrap over 5000 resamples gives $[-0.054, +0.140]$. That interval, rather than the $p$-value, is the quantitative result: conditional on the included covariates, any association between LoTSS component class and atomic hydrogen content is smaller than about $0.15$\,dex, a factor of $1.4$. We find no evidence for such an association and do not claim to have established its absence. The redshift term should be read as absorbing the FASHI selection function rather than as a physical dependence, since in a flux-limited 21\,cm survey the minimum detectable $\MHI$ rises steeply with distance \citep{2011AJ....142..170H, 2018ApJ...861...49H}.

The stellar-mass coefficient deserves comment, because its apparent insignificance is itself a selection effect. Adding the rest-frame $g-r$ color as a proxy for star formation raises it to $+0.128 \pm 0.055$, with a color coefficient of $-0.248 \pm 0.128$\,dex\,mag$^{-1}$: at fixed color, stellar mass does predict H\,{\sc i} mass, and the two dependences largely cancel when color is omitted because more massive hosts in this sample are redder and less H\,{\sc i}-rich. The component-class coefficient is unaffected by this addition.

Table~\ref{tab:regalt} lists the component-class coefficient across alternative specifications. It lies between $+0.032$ and $+0.046$\,dex across all nine specifications, every interval spanning zero, whether the $g-r$ color is included, whether extreme gas fractions, faint detections or photometric host redshifts are removed, and whether the 15 systems flagged as in-beam confused (Section~\ref{subsec:association}) are excluded. Further diagnostics are given below.

Two variables identified as potentially relevant are absent from Equation~(\ref{eq:ols}). No star formation rate is available from the LoTSS DR2 value-added catalog, and the $g-r$ color is only a proxy for one. Environment is uncontrolled, which is the more serious omission, since FAST-beam confusion is itself an environmental effect. The bound quoted above is therefore conditional on the included covariates.

\subsection{Diagnostics and alternative specifications}
In the mass-restricted sample the \texttt{M}-coded sources have a median gas fraction $\log_{10}\fHI = -0.264$ $[-0.564, -0.004]$ against $-0.098$ $[-0.157, -0.040]$ for the \texttt{S}-coded sources, an offset of $0.167$\,dex; in the 473-source strict-association sample the offset is $0.140$\,dex. The median gas fraction of the \texttt{M} class is poorly determined --- its 95 \% interval spans $0.56$\,dex against $0.12$\,dex for the \texttt{S} class --- and its interquartile range is asymmetric, so the offset is driven by a tail of gas-poor sources rather than a uniform shift.

Variance inflation factors for Equation~(\ref{eq:ols}) are $1.14$ (component class), $1.69$ ($\log\Mstar$), $2.05$ ($z$) and $2.80$ ($\log\Lfoursp$), so the individual coefficients are not compromised by collinearity. A Breusch--Pagan test gives $p = 0.54$, consistent with homoscedastic residuals, and the HC3 standard errors we quote throughout are indistinguishable from the nominal ones. The residuals are mildly leptokurtic (kurtosis $3.8$) but symmetric (skew $-0.12$). All four sources with $|\Delta v| > 300$\,km\,s$^{-1}$ are among the six photometric-redshift hosts.

\begin{deluxetable}{lccc}
\tablecaption{Component-class coefficient $b_1$ under alternative specifications, fitted on the 500-source mass-analysis sample. Standard errors are heteroscedasticity-robust (HC3).
\label{tab:regalt}}
\tablehead{\colhead{Specification} & \colhead{$N$} & \colhead{$b_1$ (dex)} &
\colhead{adj. $R^2$}}
\startdata
Primary, response $\log_{10}\MHI$      & 500 & $+0.045 \pm 0.051$ & 0.499 \\
\hline
$+\ (g-r)$ color                        & 500 & $+0.046 \pm 0.050$ & 0.501 \\
Excluding in-beam confused ($174''$)    & 485 & $+0.045 \pm 0.050$ & 0.504 \\
Excluding in-beam confused ($87''$)     & 491 & $+0.044 \pm 0.051$ & 0.501 \\
Excluding confused ($174''$) $+$ color  & 485 & $+0.046 \pm 0.050$ & 0.507 \\
Excluding $\log_{10}\fHI > 1$           & 484 & $+0.045 \pm 0.050$ & 0.525 \\
Signal-to-noise ratio $> 10$            & 470 & $+0.043 \pm 0.049$ & 0.535 \\
Separation $< 30''$                     & 306 & $+0.032 \pm 0.067$ & 0.546 \\
Excluding photometric host redshifts    & 494 & $+0.044 \pm 0.051$ & 0.504 \\
\enddata
\tablecomments{All fits use Equation~(\ref{eq:ols}) unless otherwise noted. A fit with $\log_{10}\fHI$ as the response is omitted: it is an algebraic reparameterisation of the first row, leaving $b_1$, $b_3$ and $b_4$ identical and shifting only the stellar-mass coefficient, by exactly unity, so it constitutes no independent evidence. Every 95 per cent interval spans zero.}
\end{deluxetable}

\onecolumngrid
\section{Mid-Infrared Host Colors}
\label{app:wise}
Figure~\ref{fig:wise} shows the WISE $W1-W2$ versus $W2-W3$ colors of the hosts
in the mass-restricted sample of Appendix~\ref{app:regression}. The WISE magnitudes in the value-added catalog derive from unWISE forced photometry on the DESI Legacy imaging. The catalog documentation identifies these magnitudes as AB, and the observed colors are consistent with that convention: the median $W1-W2$ is $-0.463$ (AB), equivalently $+0.18$ on the Vega system, close to the $\simeq 0.1$ typical of normal galaxies. Because the diagnostic thresholds in the literature are defined in Vega \citep{2010AJ....140.1868W, 2011ApJ...735..112J, 2014ApJ...782...90C}, we convert them using $m_{\rm AB}-m_{\rm Vega} = 2.699$, $3.339$, $5.174$ and $6.620$ for $W1$--$W4$. We exclude sources fainter than $W2 = 19.3$ or $W3 = 16.7$ (AB), approximately the AllWISE $5\sigma$ depths; this removes 34 of the 500, leaving 466. Photometric uncertainties are available for sources within the H25 parent, and every such source in this sample has a $W3$ signal-to-noise ratio above 5; the 66 that fall outside that parent, all below its 1.1\,mJy flux limit, have none.

On a consistent system the widely used $W2-W3 > 1.5$ (Vega) criterion corresponds to $W2-W3 > -0.335$ in AB, and is satisfied by 99.8 \% of S sources and 100 \% of M sources; a single source in the sample falls below it. The criterion therefore carries no information about component class here. The median colors are $W2-W3 = 3.76$ (S) and $3.86$ (M) on the Vega system, characteristic of star-forming disks rather than quiescent early-type hosts. The criterion separates mid-infrared quiescent from star-forming galaxies, and in an H\,{\sc i}-emission-selected sample it is satisfied almost universally by construction. Figure~\ref{fig:wise} therefore characterizes the host population rather than distinguishing the component classes, and the near-universal star-forming character of these hosts is a direct consequence of selecting on 21\,cm emission.

The WISE colors provide only limited constraints on AGN activity in this sample. $W2-W3$ traces warm dust and PAH emission and correlates with specific star formation rate; it is not a measure of atomic gas content, and inferring gas richness from it would in any case be circular in an H\,{\sc i}-selected sample. The $W1-W2$ colors of our hosts show no evidence for a mid-infrared-luminous accretion-dominated population: no source satisfies the criterion $W1-W2 > 0.8$ (Vega), equivalently $W1-W2 > 0.16$ on the AB system used here, and the reddest in the sample is $0.67$ (Vega) \citep{2012ApJ...753...30S, 2012MNRAS.426.3271M, 2018ApJS..234...23A}. This does not exclude jet-mode radio AGN, which are characteristically weak in the mid-infrared, but it does mean that WISE colors alone cannot establish the presence of an AGN in these systems. We therefore use them only as a description of the host population, and defer the separation of accretion-powered from star-formation-powered radio emission to Section~\ref{subsec:agn}.

\begin{figure}[t!]
\centering	
\includegraphics[angle=0,width=8.50cm]{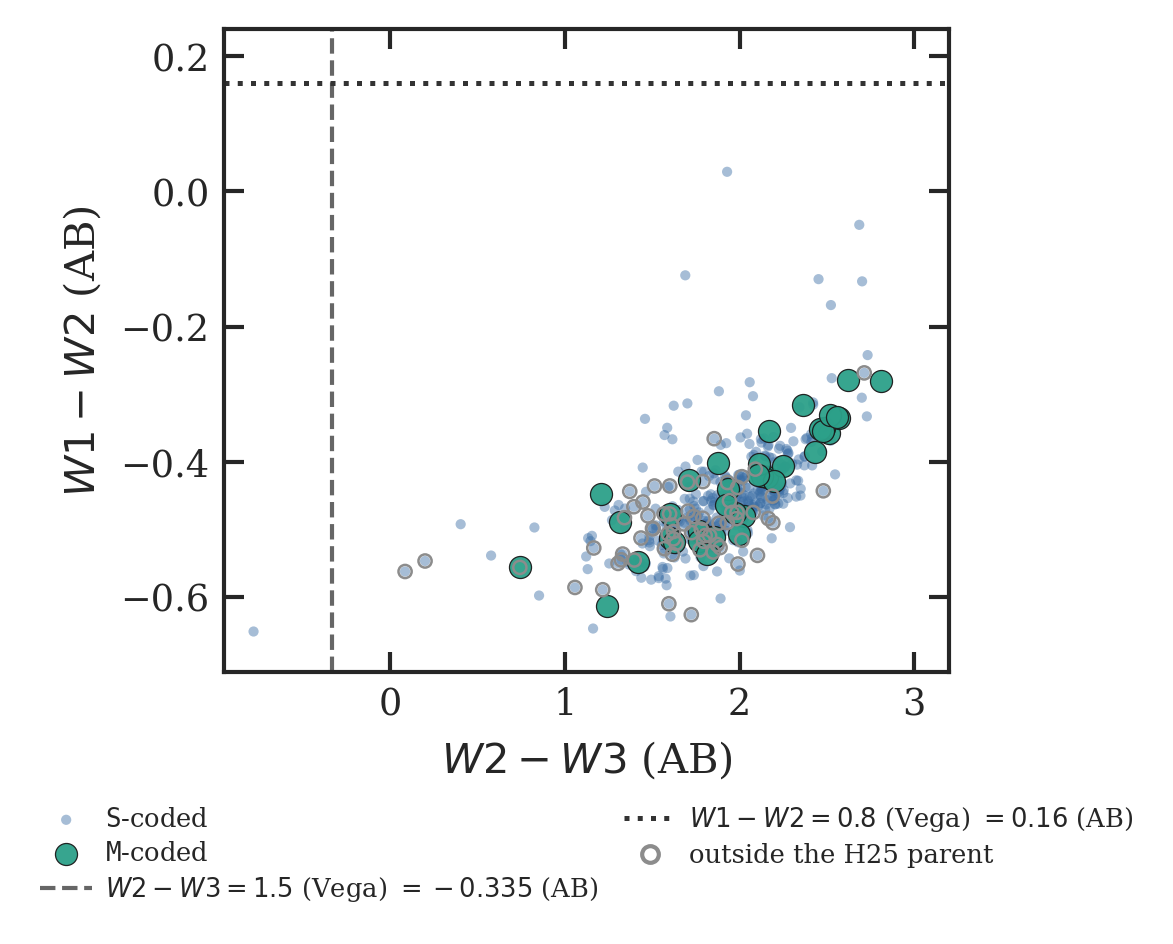}
\caption{WISE mid-infrared color--color diagram for the mass-analysis sample: \texttt{S}-coded (blue) and \texttt{M}-coded (green) LoTSS sources. Magnitudes are on the AB system; sources fainter than $W2 = 19.3$ or $W3 = 16.7$ are excluded, leaving 466 of the 500. The dashed line marks the $W2-W3 > 1.5$ (Vega) ``dusty'' criterion converted to AB ($-0.335$), satisfied by 99.8 \% of \texttt{S}-coded and 100 \% of \texttt{M}-coded sources; the dotted line marks the $W1-W2 > 0.8$ (Vega) AGN criterion, $0.16$ in AB, which no source satisfies. Grey open symbols mark the 66 sources that fall outside the H25 parent sample, all of them below its 1.1\,mJy flux limit; every source with a measured $W3$ uncertainty in this sample has a signal-to-noise ratio above 5. The hosts occupy the star-forming locus, as expected for an H\,{\sc i}-emission-selected sample.}
\label{fig:wise}
\end{figure}

\FloatBarrier
\bibliography{fastrg1_v5}{}
\bibliographystyle{aasjournal}

\end{document}